\documentclass[%
 reprint,
 amsmath,amssymb,
 aps,
]{revtex4-2}
\usepackage{graphicx}
\usepackage{dcolumn}
\usepackage{bm}

\usepackage{amsmath,amssymb,amsfonts}%
\usepackage{mathrsfs}%
\usepackage[title]{appendix}%
\usepackage{xcolor}%
\usepackage{textcomp}%
\usepackage{hyperref}

\usepackage{braket }
\usepackage{subcaption}

\begin{document}


\title{\textbf{Gain-Assisted and Dynamically Controlled Optical Bistability for Quantum Logic Gate Applications} 
}%

\author{Parkhi Bhardwaj}%
 \email{parkhi.21phz0013@iitrpr.ac.in}
\affiliation{%
Department of Physics, Indian Institute of Technology Ropar, Rupnagar, Punjab 140001, India
}%

\author{Poonam Yadav}
\affiliation{Department of Physics, Indian Institute of Technology Delhi, Hauz
Khas, New Delhi, 110016, India.
}

\author{Bodhaditya Santra}
\affiliation{%
Department of Physics, Indian Institute of Technology Delhi, Hauz
Khas, New Delhi, 110016, India.}%

\author{Shubhrangshu Dasgupta}
\affiliation{Department of Physics, Indian Institute of Technology Ropar, Rupnagar, Punjab 140001, India.
}

\date{\today}

\begin{abstract}
The propagation of a probe field in an N-type four-level cold atomic system is investigated under the influence of multiple coherent fields. Coherent control of quantum interference enables switching of the probe field between transparency and gain regimes. Subsequent analysis focuses on how the introduction of gain in the probe transition lowers the threshold for optical bistability, thereby enhancing the system’s nonlinear response at reduced input intensities. A detailed analysis of optical bistability is presented, focusing on its threshold, stability, and switching efficiency as functions of field strengths and detunings. Structured light beams, specifically Laguerre-Gaussian modes carrying orbital angular momentum, are employed to tailor the bistable characteristics. The impact of Orbital angular momentum through the topological charge and azimuthal phase is shown to significantly influence the bistable behavior. Based on these features, a theoretical scheme is proposed to realize a Controlled-NOT gate via dynamic modulation of bistability. These results highlight the potential of integrating nonlinear optical effects with structured light in cold atomic systems for implementing scalable quantum logic and advancing photonic information processing.
\end{abstract}

\maketitle


\section{Introduction}

The interaction between electromagnetic (EM) radiation and atomic systems forms the foundation of numerous phenomena in both classical and quantum optics. When an EM field induces transitions between discrete atomic energy levels, it modifies the optical response of the medium, resulting in a wide range of linear and nonlinear optical effects. At low field intensities, the medium exhibits linear behavior, where light propagation is governed primarily by refraction and absorption~\cite{r1,r2}. However, as the intensity increases, nonlinear optical phenomena emerge, giving rise to effects such as frequency conversion~\cite{r4}, self-phase modulation~\cite{r5,Schmidt:98}, and coherent photon–photon interactions.

Control of atomic properties using coherent fields gives rise to several interesting effects, including electromagnetically induced transparency (EIT)~\cite{PhysRevLett.64.1107,PhysRevLett.66.2593}, coherent population trapping (CPT)~\cite{r6}, and Autler–Townes splitting~\cite{PhysRev.100.703}. Such coherent control techniques are useful to manipulate light-matter interactions at the quantum level, facilitating a wide range of applications, e.g., slow-light propagation~\cite{r31}, optical information storage~\cite{r9}, Kerr-type optical nonlinearities~\cite{r32}, and the implementation of quantum logic gates~\cite{r34}. The presence of atomic coherence further enables tunability between transparency  and absorption regimes~\cite{r39}, offering a high degree of control over the transmission properties of the medium. In addition to these coherence-mediated effects, spontaneous and stimulated scattering processes also contribute significantly to the modification of light propagation in atomic systems~\cite{r10}.

A particularly important class of nonlinear effects, the optical bistability (OB), arises from the coherent control of atomic states. The OB has received significant attention due to its ability to support two distinct stable output states of the field for the same input conditions. This behavior results from the nonlinear interaction between the optical field and the atomic medium, governed by mechanisms such as absorption, dispersion, feedback, and atomic coherence~\cite{r13,r12}. Owing to the existence of these two stable states, OB systems are well-suited for implementing optical logic gates, which form the foundational components of digital photonic processing. Notably, Walker has demonstrated that arrays of bistable optical elements can be configured to realize parallel all-optical logic architectures~\cite{r38}. Such systems are also promising for applications in optical switching and memory, where fast, reversible control over the output state is essential.

The emergence and control of OB can be significantly enhanced by quantum interference effects such as EIT~\cite{r14,r17} and four-wave mixing (FWM)~\cite{r15,r20}, both of which serve to lower the bistability threshold and improve switching contrast. Previous theoretical and experimental studies have reported OB in various atomic configurations, including both free-space and cavity-based systems~\cite{r16,r13,r30,r22,r21,r19}. Early investigations primarily focused on relatively simple level schemes and standard field geometries. For instance, Harshawardhan and Agarwal~\cite{r28} explored OB in three-level ladder and $\Lambda$-type systems using quantum interference to control switching thresholds, while Joshi et al.~\cite{r29} studied V-type systems where OB was observed under specific detuning conditions. Kumar and Dasgupta~\cite{r27} examined a four-level ladder system with cooperative feedback, demonstrating control over OB using the strength of the applied control field. The OB has also been realized in a double-cavity N-type configuration~\cite{r23}, as well as in two-level systems with anisotropy introduced via external magnetic fields~\cite{r18}.

In recent years, OB in cold atomic ensembles has received increasing attention due to the suppression of Doppler broadening and the preservation of long-lived coherence. Cold atoms coupled to high-finesse optical resonators have exhibited clear bistable transitions, hysteresis, and dissipative phase transitions under strong coupling conditions~\cite{PhysRevA.99.013849}. These observations, often modeled using semiclassical and mean-field approaches, have advanced the understanding of cavity-enhanced optical nonlinearities and their potential applications in nonlinear quantum optics.

Beyond atomic systems, OB has also been demonstrated in various nonlinear materials~\cite{r24,r25}. However, most of these studies have employed either plane-wave or Gaussian field profiles, without accounting for the spatial complexity introduced by structured light. A particularly promising development involves the use of Laguerre-Gaussian (LG) beams, which carry orbital angular momentum (OAM). The helical phase and the ring-shaped intensity distribution of LG beams introduce new degrees of freedom, such as the topological charge (TC) and azimuthal phase, which significantly influence the optical response of nonlinear media. Recent studies have shown that higher-order LG modes can induce phase-sensitive bistability due to spatial variation of the optical field and transfer of OAM to the medium~\cite{r36,r37}.

Although previous studies on OB have primarily focused on simplified atomic configurations  — often in the context of EIT — where bistability typically emerges because of saturation-induced nonlinearity, the present work introduces several significant advancements. In particular, OB is investigated in a cold N-type atomic medium incorporating a unidirectional ring-type optical feedback configuration, wherein the nonlinearity arises predominantly from coherence-induced gain and absorption, thereby eliminating the need for high probe intensities to achieve bistability. A key contribution of this study is the integration of structured light, specifically LG beams carrying OAM, to enable spatially resolved control over the bistable response. This approach, largely unexplored in prior literature, allows for tunability through the TC and azimuthal phase associated with the LG modes, both of which are shown to significantly influence the bistability characteristics. Additionally, our study incorporates a time-dependent control field, which enables the dynamic modulation of bistable behavior and allows reversible switching between bistable and monostable regimes. Leveraging these capabilities, the study further develops a theoretical framework for the realization of an all-optical controlled-NOT (CNOT) gate, thus establishing a promising pathway for the implementation of reconfigurable quantum logic operations in coherently prepared cold atomic systems.

The article is organized as follows: Section~\ref{sec2} describes the energy-level configuration of the cold N-type atomic system and presents the dynamical equations governing its evolution under the influence of applied probe, control field, and feedback. Subsection~\ref{sub2.1} provides a comprehensive analysis of the lasing conditions, without imposing constraints on the intensities of the interacting fields. In Section~\ref{sec:ob}, the phenomenon of optical bistability is examined in detail, with particular focus on the threshold behavior, stability regions, and switching efficiency. This section also investigates the impact of critical system parameters, including the field intensities, frequency detunings, cooperative parameter \(C\), and the OAM of the structured input beams. Section~\ref{sec:cnot} introduces a theoretical framework for realizing a CNOT gate based on the dynamic modulation of bistability. Section~\ref{sec:experiment} discusses the experimental feasibility of the proposed scheme, addressing practical considerations for implementation in cold atom setups. Finally, Section~\ref{sec:conclusion} summarizes the key findings of the study and outlines their implications for future advancements in nonlinear optics, photonic logic, and quantum information processing.

\section{Description of Model}\label{sec2}
\begin{figure}
\centering
    \includegraphics[width=8cm]{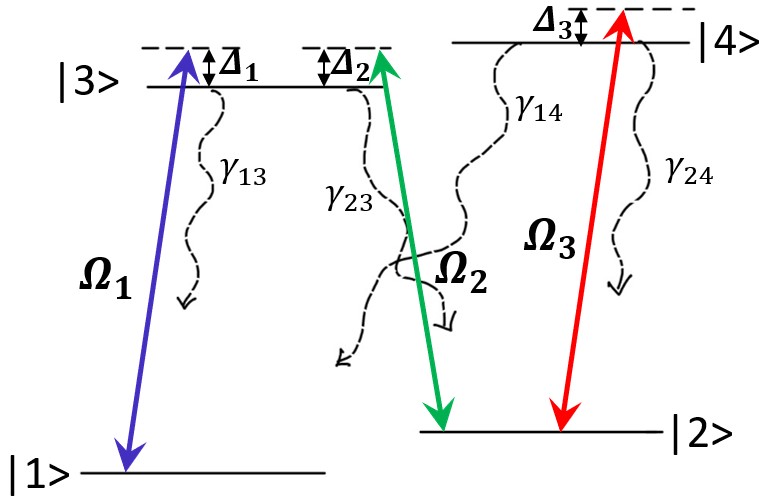}
    \caption{ Relevant energy level configuration of the  N-type four-level atomic system.}
\label{fig1}
\end{figure}

The energy level configuration of the atomic system is of the N-type, as illustrated in Fig.~\ref{fig1}. The transitions $\ket{1} \leftrightarrow \ket{3}$, $\ket{2} \leftrightarrow \ket{3}$, and $\ket{2} \leftrightarrow \ket{4}$ are electric dipole-allowed, characterized by the electric dipole moments $\vec{\mu}_{13}$, $\vec{\mu}_{32}$, and $\vec{\mu}_{42}$, respectively. All other transitions are electric dipole-forbidden.

A control field  $\vec{E}_1$ with frequency $\omega_1$ is applied to drive the transition $\ket{1} \leftrightarrow \ket{3}$ and a coupling field $\vec{E}_3$ with frequency $\omega_3$ is used to drive $\ket{2} \leftrightarrow \ket{4}$. The optical bistability of the probe field $\vec{E}_2$ with frequency $\omega_2$, coupling the states $\ket{3} \leftrightarrow \ket{2}$, is addressed in this study. 
The Rabi frequencies for the respective transitions are defined as:
\begin{equation}
\Omega_1 = \frac{\vec{\mu}_{13} \cdot \vec{E}_1}{\hbar}, \quad \Omega_3 = \frac{\vec{\mu}_{42} \cdot \vec{E}_3}{\hbar}, \quad \Omega_2 = \frac{\vec{\mu}_{32} \cdot \vec{E}_2}{\hbar}\;.
\end{equation}

The interaction between the atomic medium and applied field can be described by the following Hamiltonian under the dipole and rotating wave approximation (RWA):
\begin{align}
H &= \hbar(\Delta_2-\Delta_1)\ket{2}\bra{2} 
+ \hbar\Delta_1 \ket{3}\bra{3} \nonumber \\[6pt] 
&+ \hbar(\Delta_2-\Delta_1-\Delta_3)\ket{4}\bra{4} 
\quad - \big( \hbar\Omega_1 \ket{3}\bra{1} 
+ \hbar\Omega_2 \ket{3}\bra{2} \nonumber \\[6pt]
&\quad\;\; + \hbar\Omega_3 \ket{4}\bra{2} + h.c. \big)\; .
\label{1}
\end{align}

Here $\Delta_1 = \omega_1-\omega_{31}$, $\Delta_2= \omega_2-\omega_{32}$, and $\Delta_3=\omega_3-\omega_{42}$ are the detuning of corresponding applied fields and $\omega_{ij}$ are frequency difference between the $i$th and $j$th energy level, where $i,j \in 1, 2, 3$ and $i$ $\neq$ $j$.

The dynamical transformation of the atomic system in the presence of applied fields can be described by the Liouville equation $\dot{\rho} =-\frac{i}{\hbar}[H, \rho] + L\rho$, where $L\rho$ describes the relaxation of the atomic system by spontaneous decay. Using the Hamiltonian (\ref{1}) in the Liouville equation, the density matrix equations can be obtained as follows:

\begin{align}
\dot{\rho}_{11} &= \gamma_{13}\rho_{33} + \gamma_{14}\rho_{44} 
+ i\Omega_1^* \rho_{31} - i\Omega_1 \rho_{13}, \nonumber \\[6pt]
\dot{\rho}_{22} &= \gamma_{23}\rho_{33} + \gamma_{24}\rho_{44} 
+ i\Omega_2^* \rho_{32} - i\Omega_2 \rho_{23} 
+ i\Omega_3^* \rho_{42} - i\Omega_3 \rho_{24}, \nonumber \\[6pt]
\dot{\rho}_{33} &= -(\gamma_{13}+\gamma_{23})\rho_{33} 
+ i\Omega_1 \rho_{13} - i\Omega_1^* \rho_{31} 
+ i\Omega_2 \rho_{23} - i\Omega_2^* \rho_{32}, \nonumber \\[6pt]
\dot{\rho}_{21} &= -\!\left(\Gamma_{12}-i(\Delta_1-\Delta_2)\right)\rho_{21} 
- i\Omega_1 \rho_{23} + i\Omega_2^* \rho_{31} + i\Omega_3^* \rho_{41}, \nonumber \\[6pt]
\dot{\rho}_{31} &= -(\Gamma_{31}-i\Delta_1)\rho_{31} 
+ i\Omega_2 \rho_{21} + i\Omega_1(\rho_{11}-\rho_{33}), \label{3} \\[6pt]
\dot{\rho}_{32} &= -(\Gamma_{32}-i\Delta_2)\rho_{32} 
+ i\Omega_1 \rho_{12} - i\Omega_3 \rho_{34} 
+ i\Omega_2(\rho_{22}-\rho_{33}), \nonumber \\[6pt]
\dot{\rho}_{34} &= -\!\left(\Gamma_{34}-i(\Delta_2-\Delta_3)\right)\rho_{34} 
+ i\Omega_1 \rho_{14} - i\Omega_2 \rho_{24} - i\Omega_3 \rho_{32}, \nonumber \\[6pt]
\dot{\rho}_{41} &= -\!\left(\Gamma_{41}-i(\Delta_1-\Delta_2+\Delta_3)\right)\rho_{41} 
+ i\Omega_3 \rho_{21} - i\Omega_1 \rho_{43}, \nonumber \\[6pt]
\dot{\rho}_{42} &= -(\Gamma_{42}-i\Delta_3)\rho_{42} 
+ i\Omega_3(\rho_{22}-\rho_{44}) - i\Omega_2 \rho_{43}. \nonumber
\end{align}

The population conservation condition is given by \( \sum_{i=1}^{4} \rho_{ii} = 1 \). The spontaneous decay rate from state \( |j\rangle \) to state \( |i\rangle \) is denoted by \( \gamma_{ij} \). The coherence dephasing rate between the states \( |i\rangle \) and \( |j\rangle \) is defined as \(\Gamma_{ij} = \frac{1}{2} \sum_k (\gamma_{ki} + \gamma_{kj}) + \gamma_{\text{coll}}\),
where \( \gamma_{\text{coll}} \) represents the collisional decay rate. For our subsequent calculations, we consider \( \gamma_{13} = \gamma_{23} = \gamma_{14} = \gamma_{42} = \gamma \), and the collisional decay rate is taken as \( \gamma_{\text{coll}} = 0.001\,\gamma \).

\subsection{Switching between absorption and gain}\label{sub2.1}
As mentioned before, the fields $\Omega_1$ and $\Omega_3$ serve as the control field and the coupling field, respectively, while $\Omega_2$ acts as the probe field. Our analysis is focused on the transition between the states $\ket{3}$ and $\ket{2}$ since this transition is associated with the probe field. The susceptibility $\chi(\omega)$ is expressed as $\chi(\omega)=\frac{\mathcal{N}|\mu_{32}|^2}{\hbar}\rho_{32}$, 
where \(\mathcal{N}\) is the number density of the medium
and $\rho_{32}$ denotes the coherence between the states $\ket{3}$ and $\ket{2}$.
The imaginary part of the susceptibility ($\text{Im}(\chi)$), is proportional to the absorption coefficient of the medium. When $\text{Im}(\chi)$ is positive (negative), it indicates that the field experiences absorption (gain or amplification) as it propagates through the medium. 

To obtain $\chi(\omega)$, we solve the Eqs. (\ref{3}) in the steady state for $\rho_{32}$. In this calculation, no approximations are made regarding the strength of any applied field, ensuring a general solution.
\begin{figure*}
\centering
\begin{subfigure}{.49\linewidth}
    \includegraphics[width=8cm]{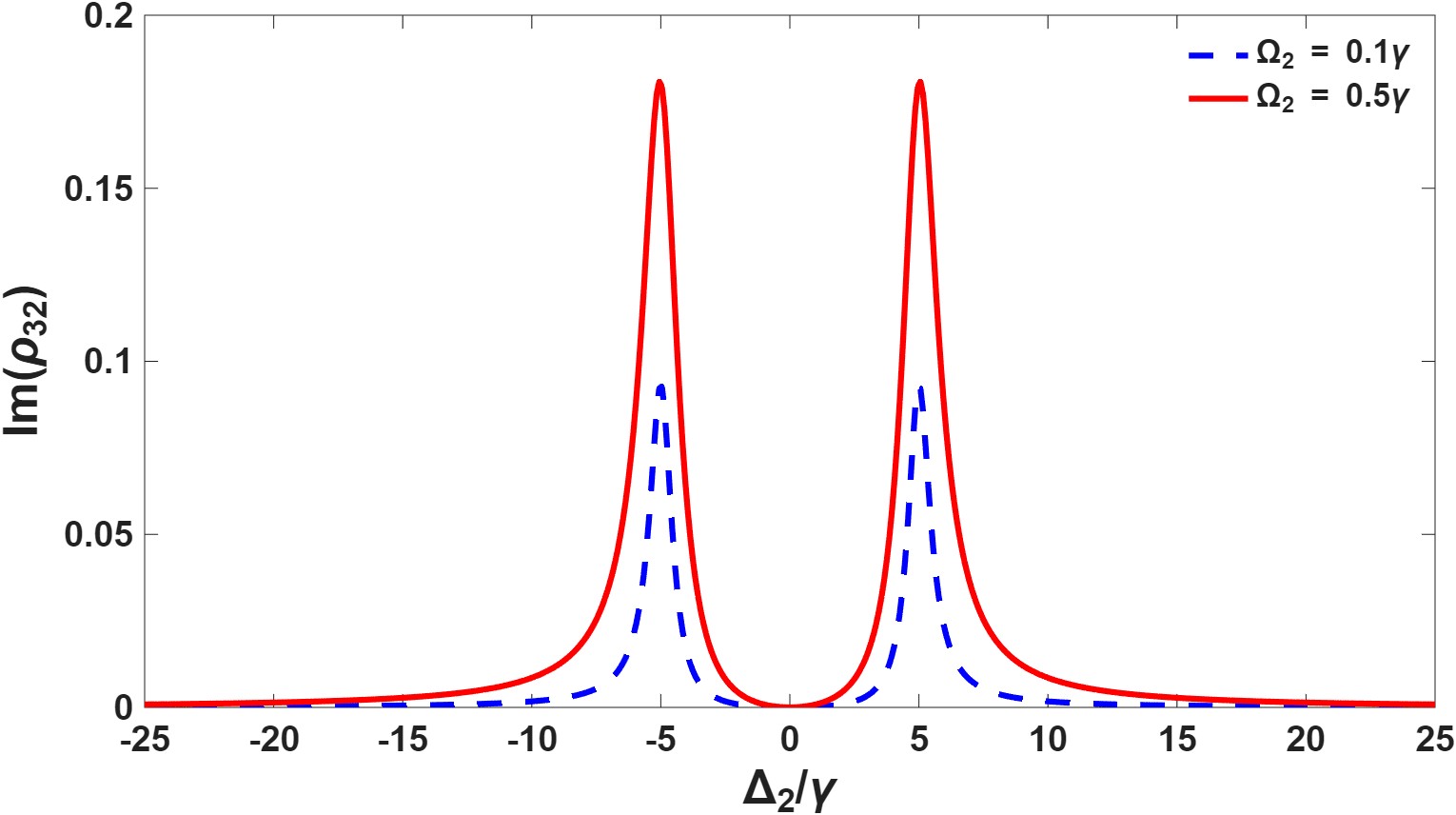}
    \caption{}
    \label{2a}
\end{subfigure}
\hfill
\begin{subfigure}{.5\linewidth}
    \includegraphics[width= 8cm]{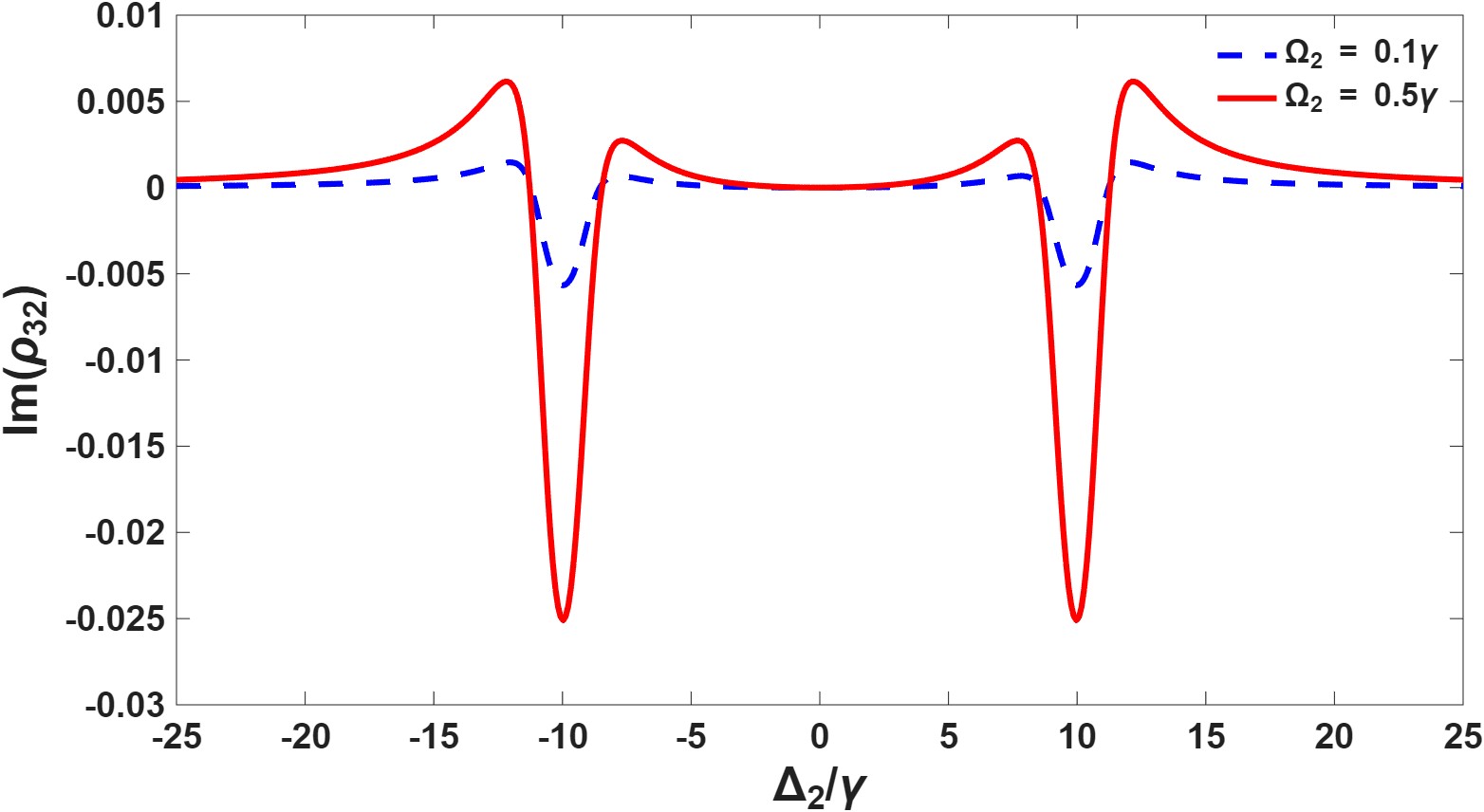}
    \caption{ }
    \label{2b}
\end{subfigure}
\caption{  $\text{Im}(\rho_{32})$ as a function of probe detuning $\Delta_2/\gamma$, for the probe field Rabi frequencies $\Omega_2 =0.1\gamma$ (blue dashed) and $\Omega_2=0.5\gamma$ (red solid). We have chosen (a)  $\Omega_1 = 5\gamma$ and $\Omega_3 = 0$, (b) $\Omega_1 = \gamma$, $\Omega_3 = 10\gamma$. All other parameters are $\Delta_1 = \Delta_3 = 0$.}
\end{figure*}
 In Figs.~\ref{2a} and \ref{2b}, we display the absorption spectrum in the absence and in the presence of the coupling field $\Omega_3$, respectively.  When the field $\Omega_3 = 0$, the system behaves as a conventional $\Lambda$-type EIT medium. For both values of the probe field strength ($\Omega_2 = 0.1\gamma$ and $0.5\gamma$), the absorption profile shows a narrow transparency window at $\Delta_2 = 0$.  This window becomes increasingly sharp and pronounced with stronger control field $\Omega_1$, a hallmark of destructive quantum interference that suppresses absorption at resonance. At $\Delta_2 = \pm 5\gamma$, prominent absorption peaks are observed, corresponding to $\text{Im}(\rho_{32}) > 0$. These peaks can be understood through the dressed-state picture, wherein the control field $\Omega_1$ splits the bare atomic states into two dressed states. When the frequency of the probe field becomes resonant with these dressed states, absorption occurs at detunings approximately equal to $\Delta_2 = \pm \Omega_1$, thus explaining the symmetric absorptive features in the spectrum.

In Fig.~\ref{2b}, the absorption spectra for a strong coupling field $\Omega_3 = 10\gamma$ is displayed.  This field couples the states $\ket{2} \leftrightarrow \ket{4}$ and thereby realizes a complete N-type four-level configuration. This additional interaction significantly alters the coherence dynamics of the medium. As shown in the Fig.\ref{2b}, the $\text{Im}(\rho_{32})$ becomes negative at detunings $\Delta_2 = \pm \Omega_1$, in the vicinity of the central transparency window, indicating the emergence of gain for the probe field. This gain arises from additional quantum interference pathways introduced by the $\Omega_3$ field. The underlying mechanism can be understood using the dressed state framework, where the coupling field $\Omega_3$ leads to a population inversion among the dressed states formed by the control and coupling fields, without requiring population inversion in the bare basis. When the probe field is resonant with these dressed states, stimulated emission dominates, resulting in amplification rather than absorption.

Such switching from absorption to gain through coherent control mechanisms demonstrates the crucial role of the coupling field in tuning the optical response of the atomic system. The interplay between the probe, control, and coupling fields enables precise manipulation of light–matter interactions, which is of significant interest for applications in all-optical switching.

\section{Optical bistability}\label{sec:ob}
\begin{figure*}
    \centering
    \includegraphics[width=15 cm]{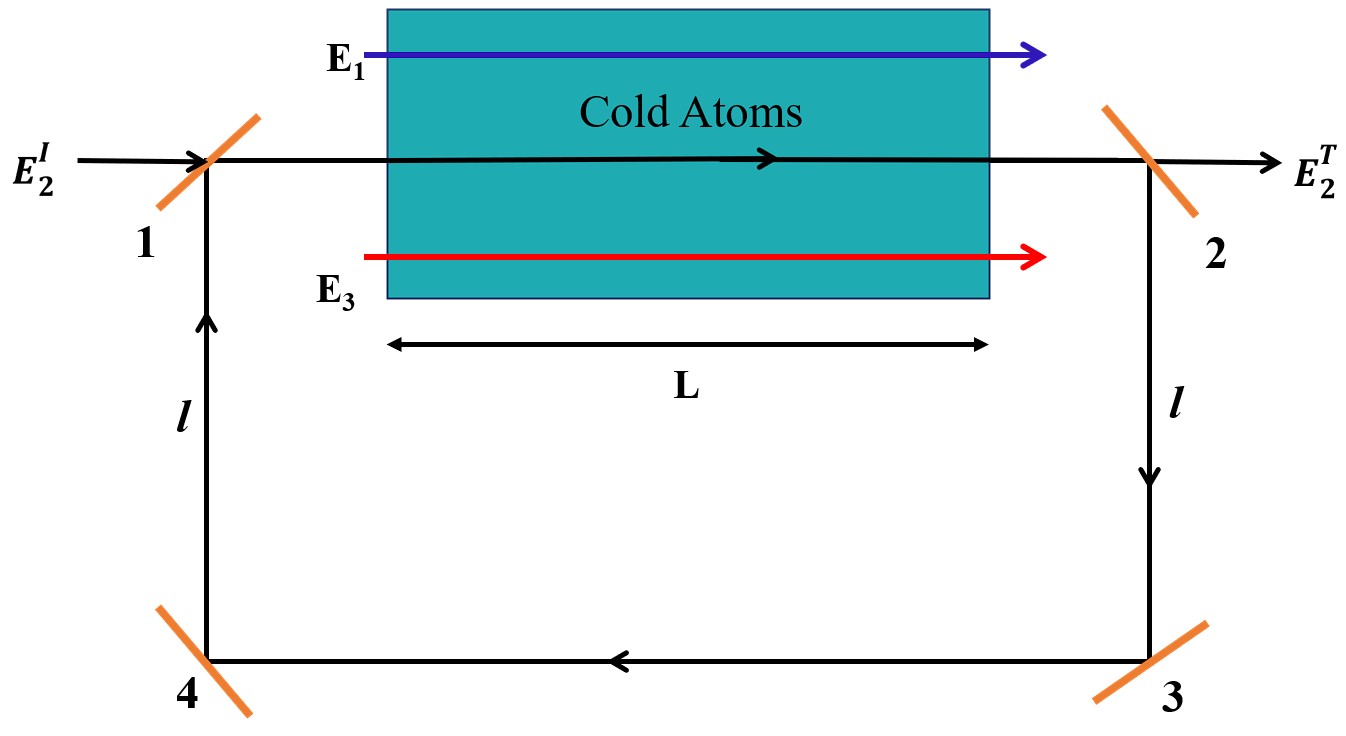}
    \caption{Experimental setup for optical bistability: Three fields, \(E_1\), \(E_2\), and \(E_3\) interact with a cold atomic medium of length \(L\), where the probe field \(E_2\) circulates within a unidirectional ring cavitywith the mirrors 3 and 4 as perfect reflectors,while the mirrors 1 and 2 obey R+T = 1.} 
     \label{ob}
\end{figure*}

The system under investigation consists of the ensemble of \(\mathcal{N}\) atoms of N-type energy level configuration placed in a unidirectional cavity as shown in Fig. \ref{ob}. In this figure, mirrors 1 and 2 are partially reflectors with the reflection coefficient $R = 1 - T$, where $T$ is the transmission coefficient, while mirrors 3 and 4 are fully reflectors with $R = 1$ and $T=0$. We apply three fields to the atomic ensemble: the control, the probe, and the coupling fields. 
The probe field with an initial amplitude $E_2^I$ partially transmits into the cavity through the mirror 1. The fields $E_1$ and $E_3$ are used to alter the optical properties of the ensemble and do not circulate in the cavity. The dynamic response of the ensemble to the probe field $E_2$ is given by the Maxwell equation, which can be written in the slowly varying envelope approximation as follows:
\begin{equation}
      \frac{1}c{}\frac{\partial E_2}{\partial t} +\frac{\partial E_2}{\partial z} = \frac{i\omega_2}{2\epsilon_oc}P(\omega_2) \;. 
      \label{6}
\end{equation}
Here, \(\epsilon_o\) is the permittivity in free space, $c$ is the speed of light in vacuum, and $P(\omega_2) = \mathcal{N}\mu_{32}\rho_{32}$ is the induced polarization of the probe field. The field $E_2^T$ is the transmitted field from mirror 2. So for a perfectly tuned cavity, the boundary conditions on the incident field and the transmitted field are as follows: 
\begin{eqnarray}
    E_2^T(t) &=& \sqrt{T}E_2(L,t), \nonumber\\
    E_2(0,t) &= &\sqrt{T} E_2^I(t) + R E_2(L,t-\Delta\tau)\;,
    \label{7}
\end{eqnarray}
where \( E_2(0, t) \) denotes the probe field at the entrance of the medium, and \( E_2(L, t) \) represents the field after it has propagated through the sample of length \( L \). The round-trip time delay between the mirrors is given by \( \Delta\tau  = \frac{2l + L}{c} \), which accounts for the time taken by the light to travel from mirror 2 to mirror 1, $l$ being the distance between the mirror 2 and the mirror 3 (and also between the mirror 4 and the mirror 1). In the steady state, the above boundary
conditions take the following form:
\begin{equation}
    E_2^T= \sqrt{T}E_2(L),
    E_2(0) = \sqrt{T} E_2^I + R E_2(L)\;.
\label{8}
\end{equation}
 In the mean-field limit and using Eq. (\ref{8}), we have the following relation between the input and the output field:
\begin{align}
y &= 2x - iC\,\rho_{32}, \nonumber \\[6pt]
y &= \frac{\mu_{32} E_2^{I}}{2\hbar\sqrt{T}},
x = \frac{\mu_{32} E_2^{T}}{2\hbar\sqrt{T}}.
 \label{10}
\end{align}

where $y$ and $x$ are the normalized input and output fields, respectively, and $C$ is the cooperativity parameter of the atomic system in the ring cavity, given by $C=\mathcal{N}\omega_2L|\mu_{32}|^2/2 \hbar\gamma \epsilon_ocT$. We next explore the stability characteristics of the branches of the expected bistability, arising out of Eq. (\ref{10}), which is inherently nonlinear. This bistability can be characterized by two quantifiers as discussed below. 

\subsection{ Stability and switching Analysis of Optical Bistable States:} 

\begin{figure}[ht]
\centering
  \includegraphics[width=8cm]{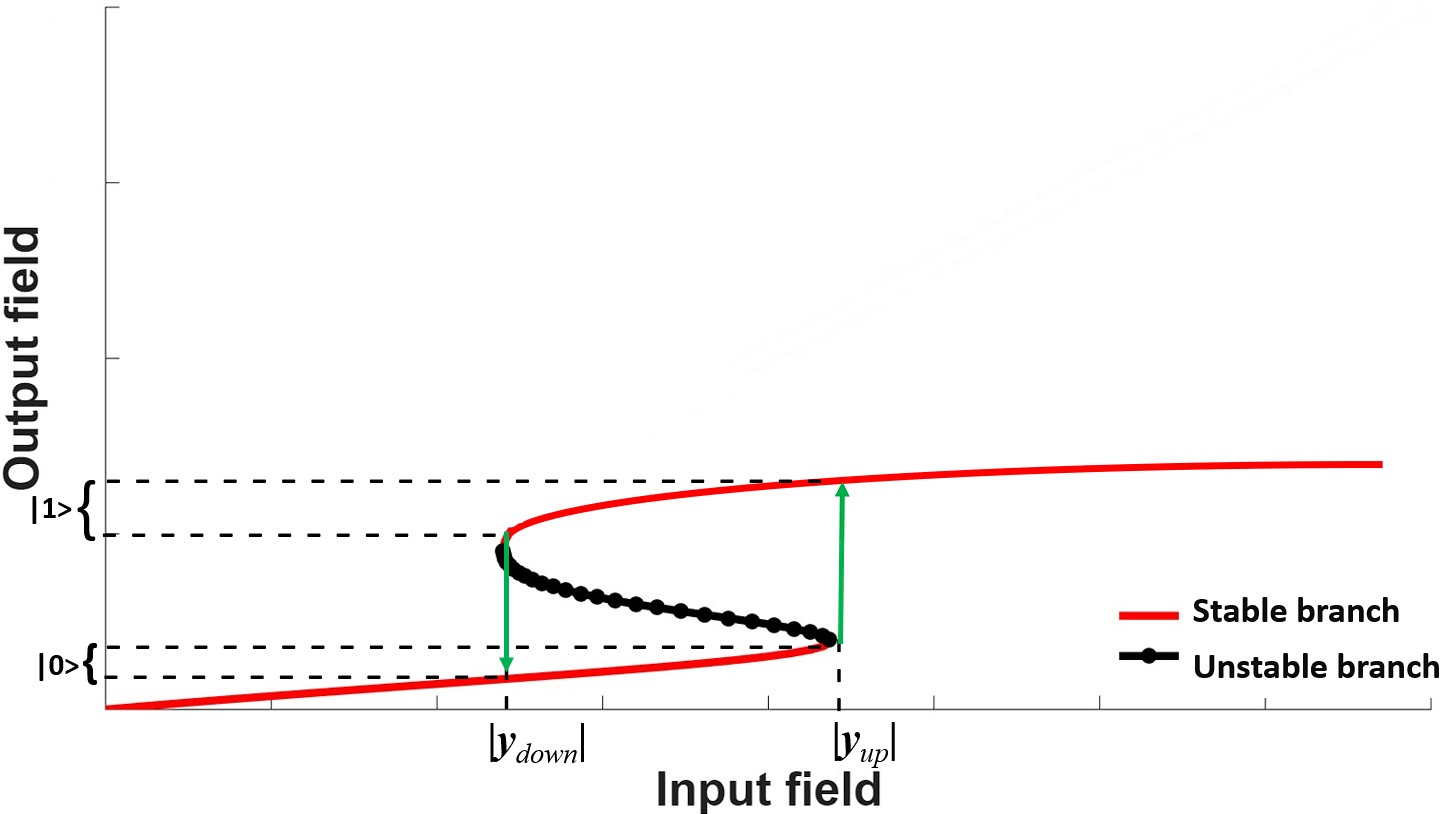}
    \caption{Hysteresis curve of optical bistability}
    \label{8a}
\end{figure}
A typical bistable behavior is characterized by a hysteresis-like loop (an S-shaped curve), with two turning points: the upper threshold (switching from low to high state) and the lower threshold (switching from high to low state) as shown in Fig.\ref{8a}. The horizontal distance between the upper and lower thresholds in the input field axis represents the length of the hysteresis loop: $L_{hys} = y_{up}-y_{down}$. 
Here, the $y_{up}$ ($y_{down}$) is the upper (lower) threshold, i.e., the input field at which the system switches from the low-output (high-output) stable state to the high-output (low-output) stable state.
This length indicates the range of input fields where the system maintains bistable behavior. A longer hysteresis loop ensures data stability, as the stored state (high or low) is less prone to accidental switching, while a shorter loop enables faster switching but may reduce reliability under noisy conditions.

In OB, the switching refers to the transitions between the two stable output states (low and high field strength) as the input field is varied. This switching behavior is directly linked to the slope of the OB curve. The OB curve has three regions: two stable regions (upper and lower branches) with a smaller slope (nearly flat) than the unstable region (middle branch) with a steep negative slope. The system switches from the lower stable branch to the upper stable branch at a critical positive slope of the lower branch, beyond which the system can no longer remain stable on the lower branch. The derivative $S=dx/dy$ can be a good marker of such critical points.
At the switching points, this slope becomes large, indicating a rapid change in $x$ with a small variation in $y$. Switch-up occurs when the 
$S$ reaches a critical positive value, and the lower branch becomes unstable. Similarly, 
 switch-down occurs when the $S$ reaches a critical negative value and the upper branch becomes unstable. Rapid changes in output enable the system to act as an optical switch. While a good storage device is marked by a large $L_{hys}$ and a small $S$, a good switch is marked by a large $S$ and a small $L_{hys}$.

\section{Numerical Results of Optical Bistability Characteristics}
 \subsection{Control of the Optical Bistability by Field strengths}\label{num ob}
\begin{figure*}
\centering
\begin{subfigure}{.49\linewidth}
    \includegraphics[width= 8cm]{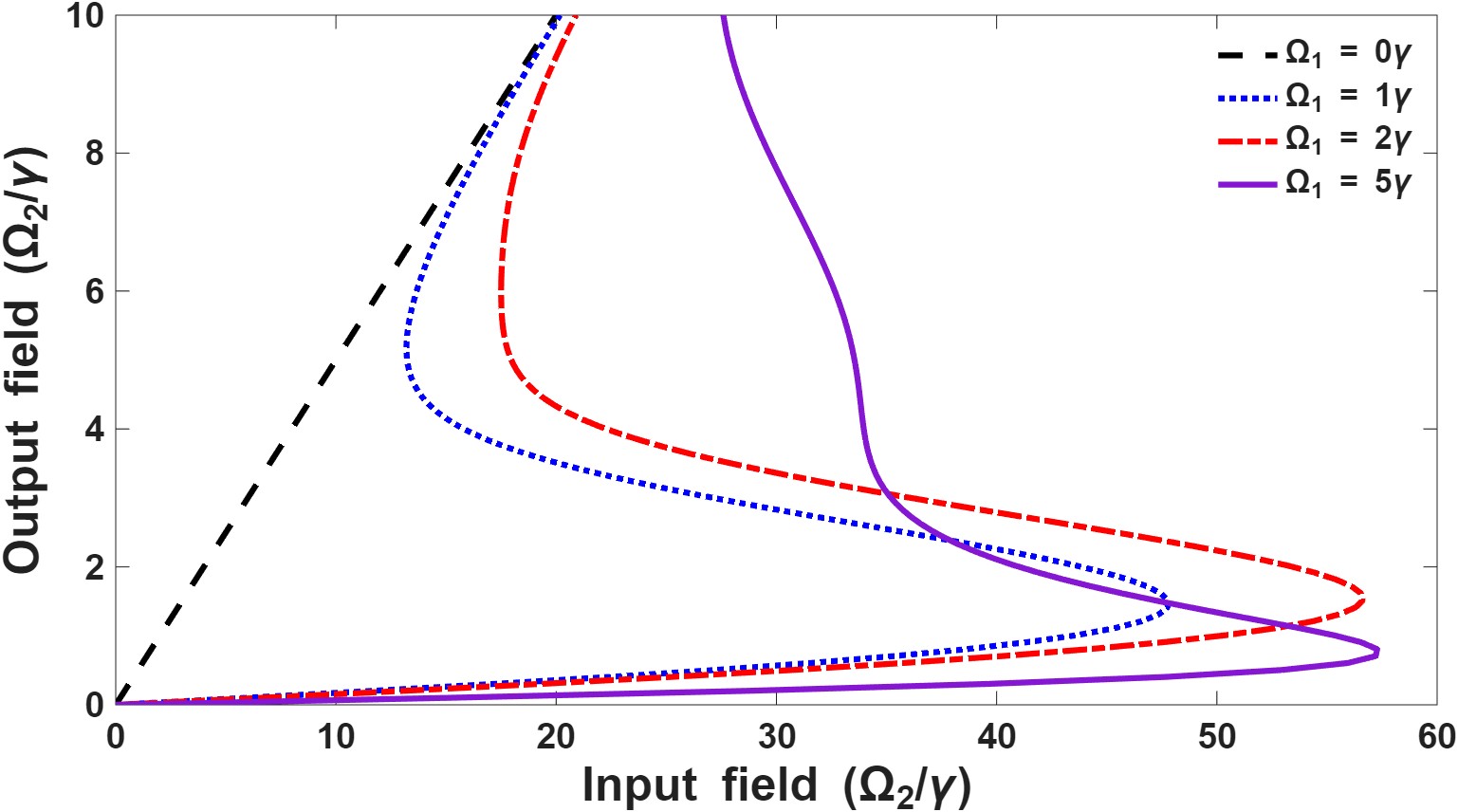}
    \caption{ }
    \label{4a}
\end{subfigure}
\hfill
\begin{subfigure}{.49\linewidth}
    \includegraphics[width=8cm]{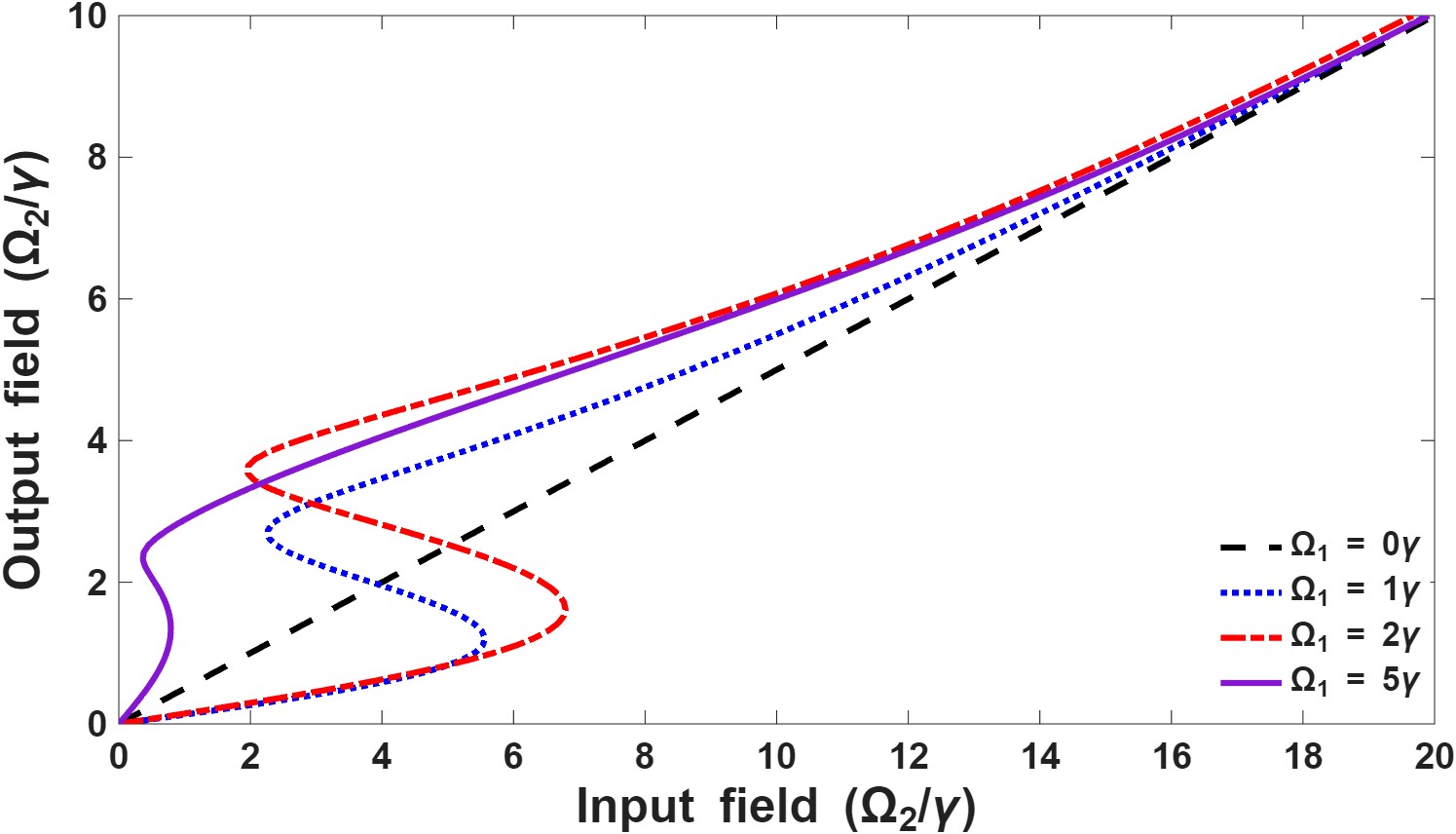}
    \caption{}
    \label{4b}
\end{subfigure}

\hfill
\begin{subfigure}{1\linewidth}
    \centering
    \includegraphics[width= 8cm]{ 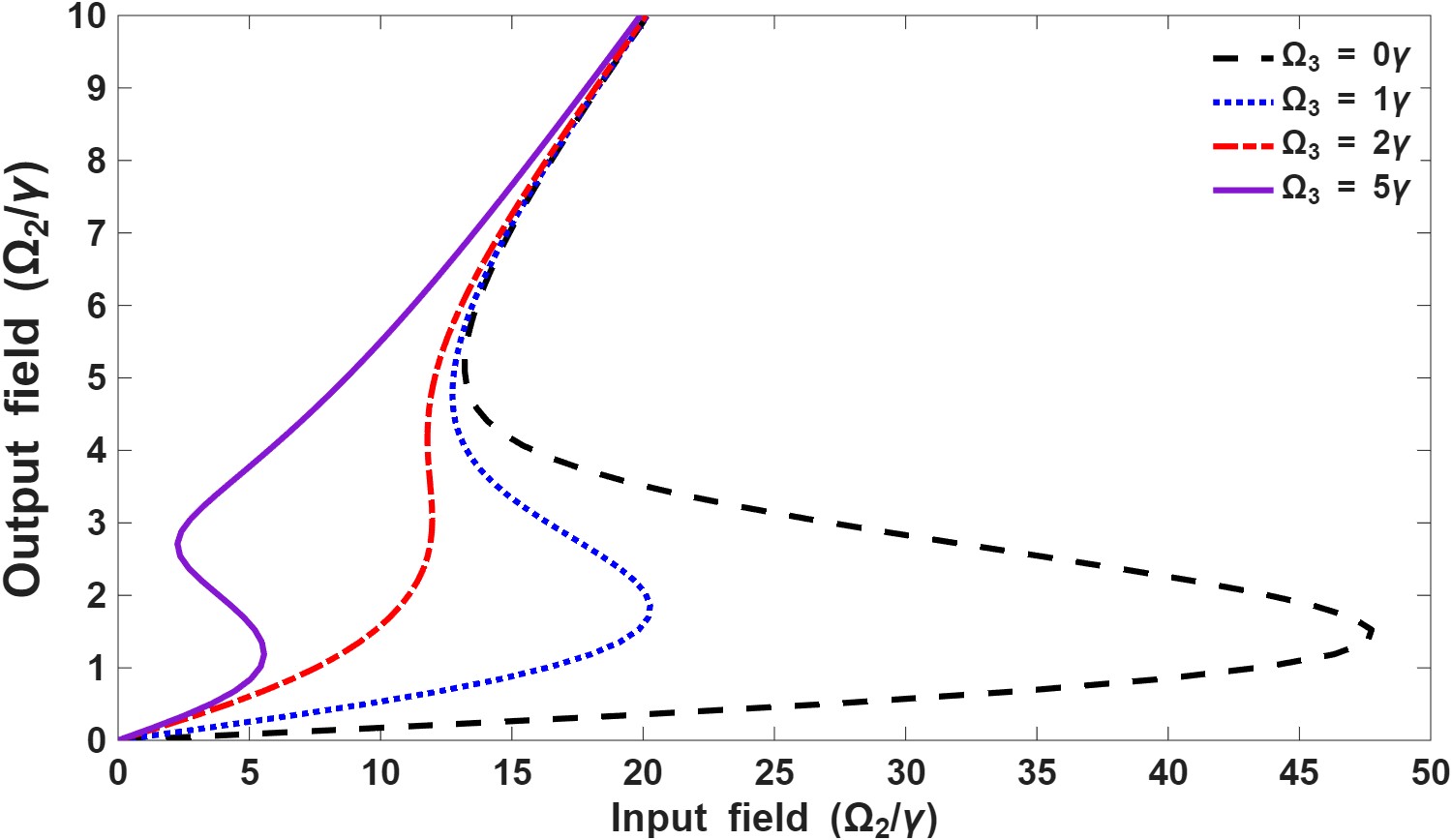}
    \caption{ }
    \label{4c}
\end{subfigure}
\hfill
\caption{Variation of OB for $\Omega_1=0$ (black, dashed), $\gamma$ (blue, dotted), 2$\gamma$ (red, dash-dotted), and 5$\gamma$ (violet, solid), for (a)  $\Omega_3 = 0$ and (b) $\Omega_3 = 5\gamma$. In subfigure (c), we have shown the results for $\Omega_3 =0$ (black, dashed), $\gamma$ (blue, dotted), 2$\gamma$ (red, dash-dotted), and 5$\gamma$ (violet, solid), for $\Omega_1 = \gamma$. The other parameters are $\Delta_2 = 5\gamma$, \(\Delta_1\)=\(\Delta_3=0\), and $C = 300$.}
\end{figure*}

Next, we present our results of the OB. 
From Eq.~\ref{10}, it is evident that \(x\) and \(y\) represent the input and output fields, respectively. Throughout this work, we plot the field strengths, given by the moduli \(|x|\) and \(|y|\). The bistability regime is determined by several key parameters, including the amplitudes of the applied fields.
 In the Figs. \ref{4a},\ref{4b}, and \ref{4c}, we show how the bistability regions vary with changes in the field Rabi frequencies $\Omega_1$ and $\Omega_3$, respectively, for a fixed atomic detuning \( \Delta_2 = 5\gamma \) and the cooperation parameter \( C = 300 \). The curves in Fig.~\ref{4a} are plotted when the coupling field $\Omega_3$ is absent. We observe that bistability does not occur if the control field $\Omega_1$ is kept switched off. The OB curves start appearing for nonzero $\Omega_1$ and as \( \Omega_1 \) increases, the OB threshold shifts toward higher input field strengths. This effect can be attributed to enhanced absorption of the probe field when its amplitude increases. Note that the usual EIT is a linear effect in the weak probe field limit, and corresponds to transparency at resonance. 

We show a similar set of curves in Fig.~\ref{4b}, in the presence of the coupling field ($\Omega_3\ne 0$). In this case, the bistability is absent whenever the control field $\Omega_1$ is kept switched off. However, the OB response can be obtained for nonzero values of $\Omega_1$. Interestingly, contrary to what is seen in the absence of the coupling field (Fig. \ref{4a}), the OB threshold gets reduced with increasing \( \Omega_1 \). This is because the coupling field introduces gain and enhances the medium’s nonlinear response at low intensities.

However, the trend in the change of the OB is quite different when one changes $\Omega_3$, by maintaining a nonzero $\Omega_1$. We show this trend in Fig.~\ref{4c}. We find that the OB exists even when the coupling field $\Omega_3$ is absent. This situation corresponds to the nonlinearity exhibited in a usual $\Lambda$ configuration for a strong probe field. This effect also underlines the importance of the control field $\Omega_1$ in obtaining the OB.  Increasing \( \Omega_3 \), however, reduces the OB threshold by introducing gain and strengthening the nonlinear interaction. At a critical value of \( \Omega_3\sim 2\gamma \), the bistable behavior vanishes. By further increasing $\Omega_3$, the OB features are regained, however, with a very low threshold. 
These observations underscore the importance of carefully tuning both \( \Omega_1 \) and \( \Omega_3 \) to achieve a low switching threshold while preserving stable bistable operation.

 \subsection{Control of the Optical Bistability by detunings}\label{num ob}
\begin{figure*}
\centering
\begin{subfigure}{.49\linewidth}
    \includegraphics[width=8cm]{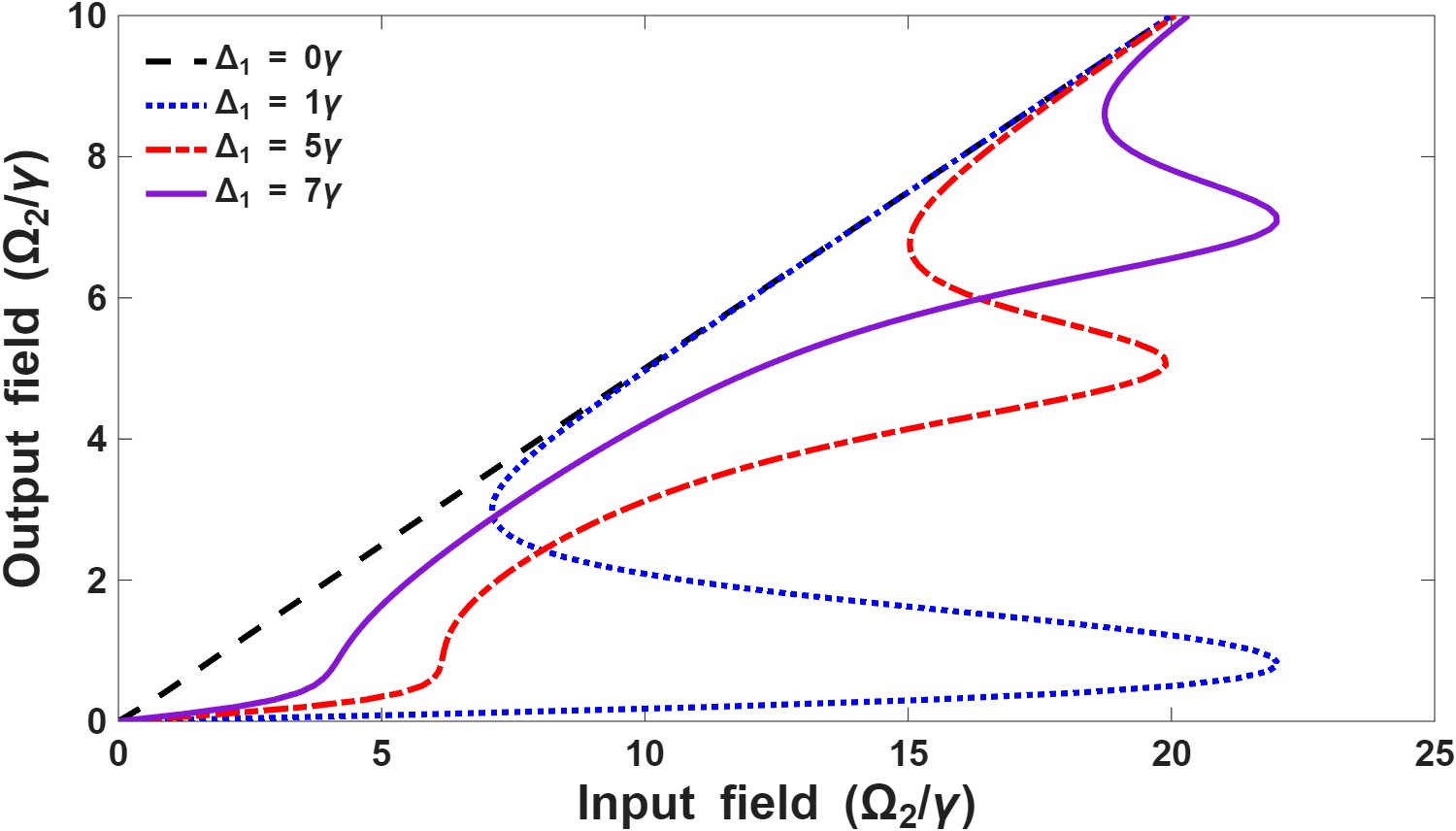}
    \caption{}
    \label{5a}
\end{subfigure}
\hfill
\begin{subfigure}{.5\linewidth}
    \includegraphics[width= 8cm]{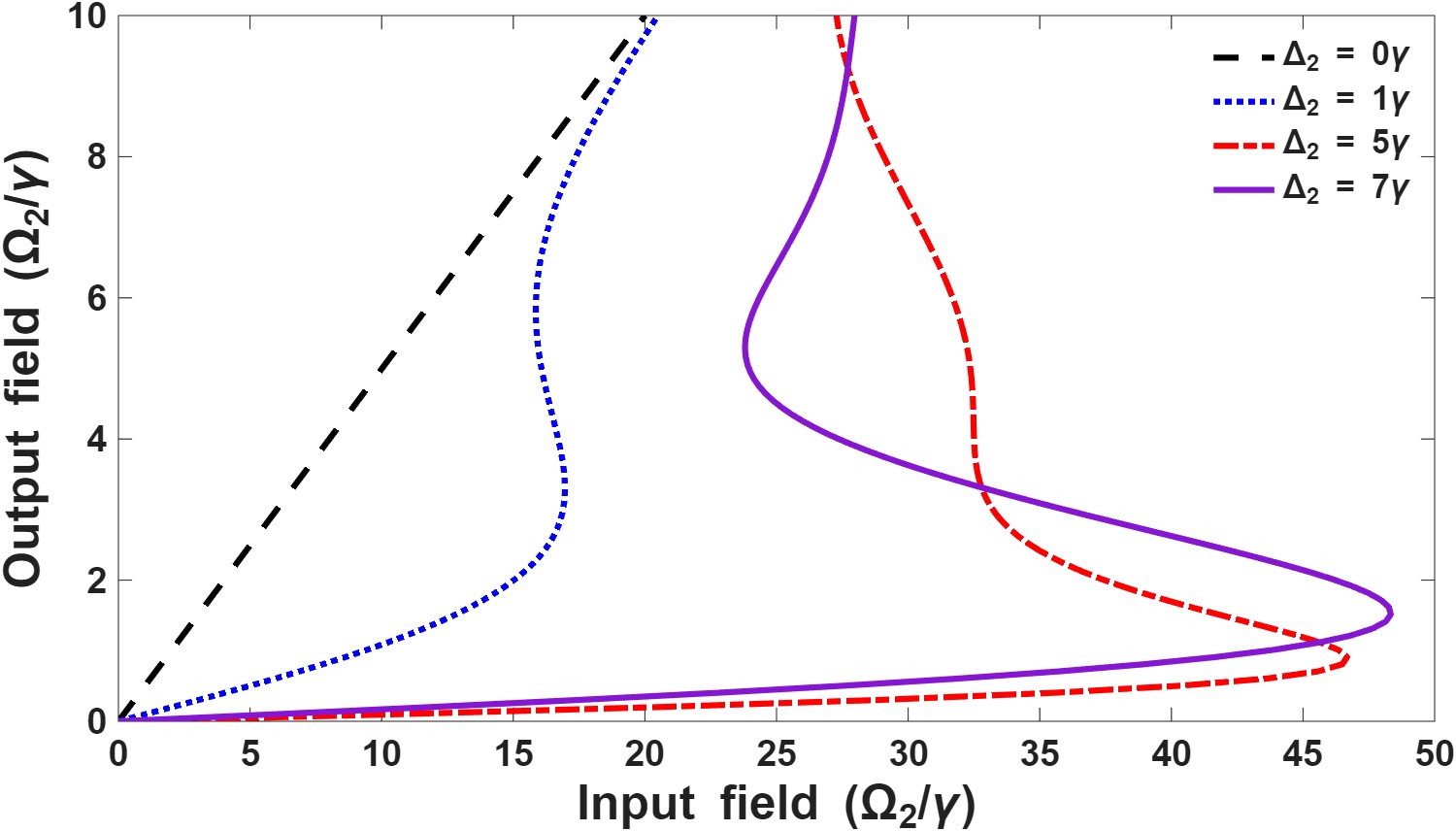}
    \caption{ }
    \label{5b}
\end{subfigure}
\hfill
\begin{subfigure}{.49\linewidth}
    \includegraphics[width= 8cm]{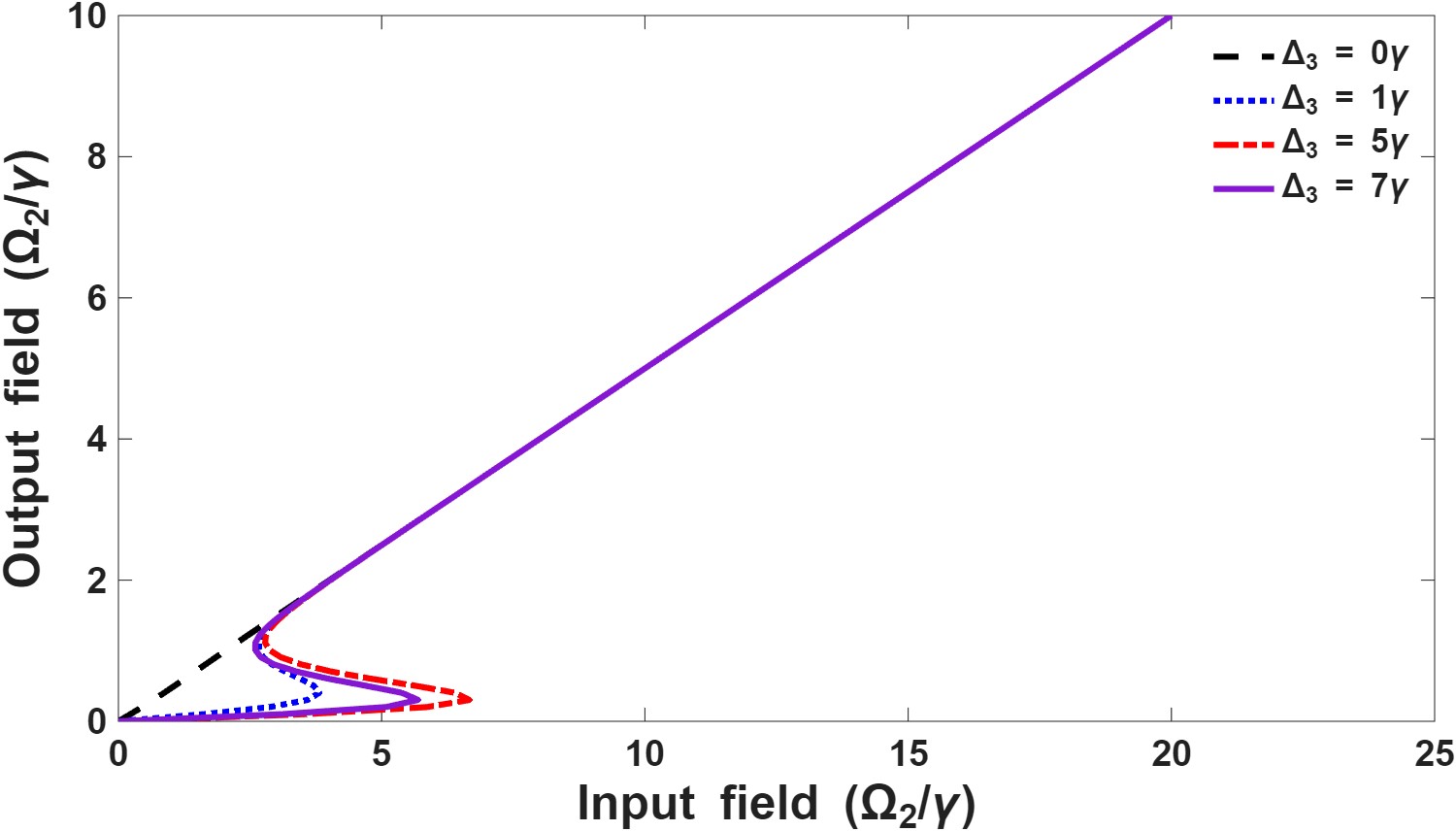}
    \caption{ }
    \label{5c}
\end{subfigure}
\hfill
\begin{subfigure}{.49\linewidth}
    \includegraphics[width= 8cm]{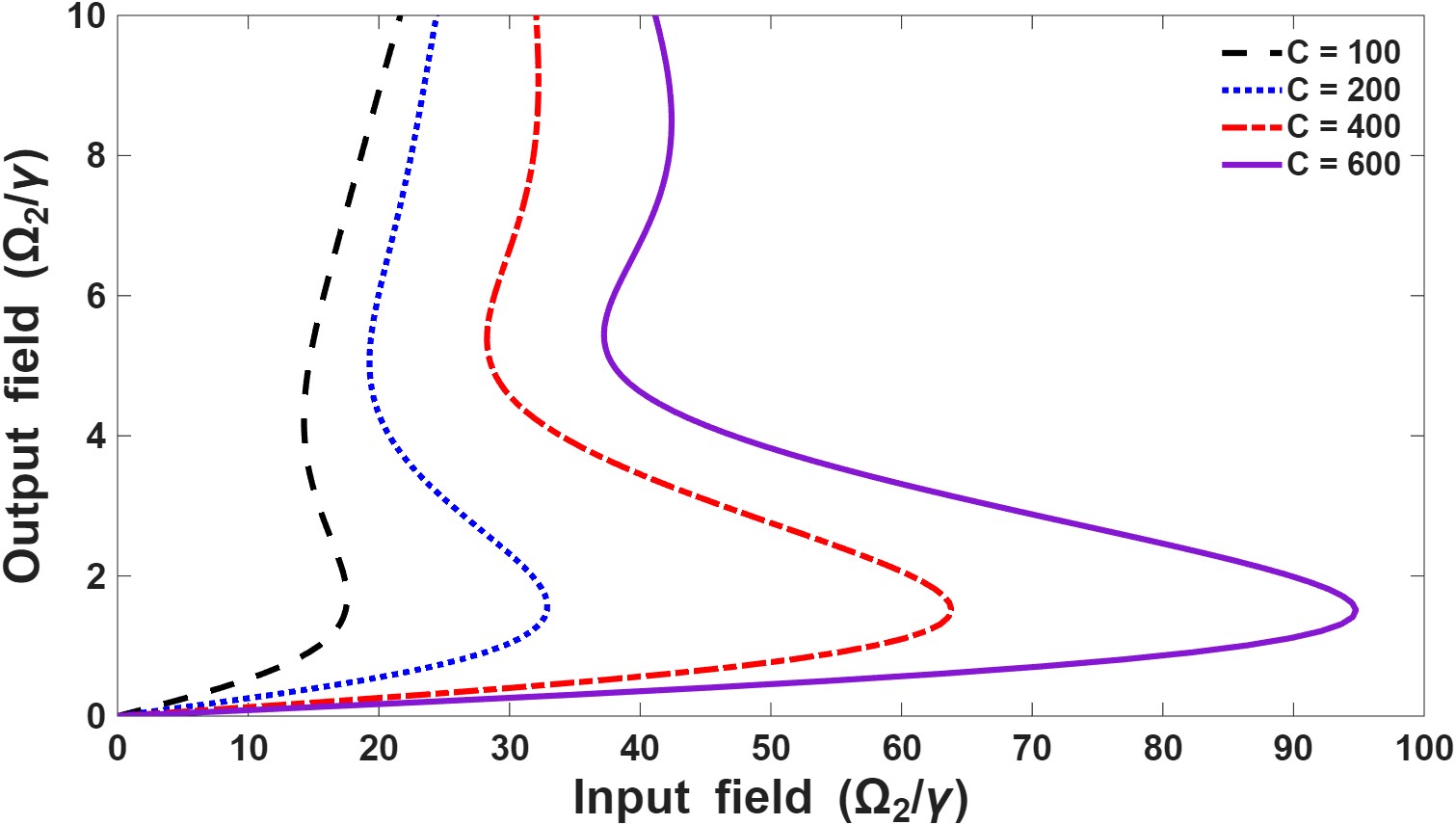}
    \caption{ }
    \label{5d}
\end{subfigure}
\caption{Variation of OB, for (a) $\Delta_1=0$ (black, dashed), $\gamma$ (blue, dotted), 5$\gamma$ (red, dash-dotted), and 7$\gamma$ (violet, solid) when $\Omega_1=0.5\gamma$, $\Omega_3=0.5\gamma$, and $C=300$, (b) $\Delta_2=0$ (black, dashed), $\gamma$ (blue, dotted), 5$\gamma$ (red, dash-dotted), and 7$\gamma$ (violet, solid) when $\Omega_1=5\gamma$, $\Omega_3=0.5\gamma$, and $C=300$, (c) $\Delta_3=0$ (black, dashed), $\gamma$ (blue, dotted), 5$\gamma$ (red, dash-dotted), and 7$\gamma$ (violet, solid) when $\Omega_1=0.5\gamma$, $\Omega_3=0.5\gamma$, and $C=300$, and (d) $C=$ 100 (black, dashed), 200 (blue, dotted), 400 (red, dash-dotted), and 600 (violet, solid) when $\Omega_1=5\gamma$, $\Omega_3=0.5\gamma$, $\Delta_1=\Delta_3=0$, and $\Delta_2$=7$\gamma$.  }
\end{figure*}

Next, we investigate the influence of varying the detunings of the three input fields on the bistability behavior of the system. The corresponding analysis is illustrated in Figs.~\ref{5a}, \ref{5b}, and \ref{5c}, which present the optical bistability curves for different values of the detunings \( \Delta_1 \), \( \Delta_2 \), and \( \Delta_3 \), respectively.

In Fig.~\ref{5a}, we analyze the effect of changing the detuning \( \Delta_1 \) of the control field $\Omega_1$. For zero detuning (\( \Delta_1 = 0 \)), the field \( E_2 \) does not exhibit any bistable behavior. However, when \( \Delta_1 \) is increased to \( 1\gamma \), a single bistability region emerges. With further increase in detuning to \( 5\gamma \) and \( 7\gamma \), two distinct bistability regions are observed. This indicates the presence of bistability at both low and high input field strength, suggesting multistable behavior of the system at higher detuning values.

A similar trend is observed in Fig.~\ref{5b}, where the detuning \( \Delta_2 \) of the probe field $\Omega_2$ is varied. At lower values of \( \Delta_2 \), a single bistability region is present. As \( \Delta_2 \) increases to \( 5\gamma \) and \( 7\gamma \), the system transitions into a regime showing multiple bistability regions. In particular, the switching characteristics and the stability of the bistable branches improve for larger $\Delta_2$. 

In Fig.~\ref{5c}, the impact of varying the detuning \( \Delta_3 \) of the coupling field $\Omega_3$ is examined. At \( \Delta_3 = 0 \), the system does not show any optical bistability. As the detuning increases to \( 1\gamma \), \( 5\gamma \), and \( 7\gamma \), a clear bistable region is formed. Interestingly, while the width of the bistable region decreases with increasing \( \Delta_3 \), the bistability threshold shifts to lower input field strengths. This behavior indicates a strengthening of the system’s nonlinearity with larger \( \Delta_3 \) values.

The detuning parameters \( \Delta_1 \), \( \Delta_2 \), and \( \Delta_3 \) play a pivotal role in controlling the optical bistability and multistability characteristics of the system. While lower detuning values tend to support the presence of a single bistability region, higher detunings lead to the emergence of multiple bistability and multistability regimes. This highlights the system's sensitivity to detuning. 

Interestingly, with increasing cooperativity $C$, the bistable behavior becomes more pronounced. We display this effect in the Fig. \ref{5d}.  For lower values of $C\sim  100, 200$, the bistability region is narrower, with a smaller separation between the bistability points on the curve.
For higher values of $C \sim 400, 600$, both bistable and multistable regions are observed, which broaden with increasing $C$. The system exhibits a larger difference between the lower and upper states. Additionally, the switching efficiency and stability of OB can be enhanced under these conditions.
     
\subsection{Control of the Optical Bistability by OAM}\label{oam ob}
Now, we will focus on the effect of the OAM of the fields on the OB. In this regard, we consider the control and the coupling field, each as a superposition of two LG beams with opposite OAM quantum numbers (i.e., the TC). Their amplitudes can thus be written as $E_1 = |A_1|(e^{il_1\phi}+e^{-il_1\phi})$ and $E_3 =|A_3|(e^{il_3\phi}+e^{-il_3\phi})$, respectively.
where $A_1$ and $A_3$ are the amplitudes of these fields at the beam waist $r=r_0$, $r$ denoting the radial distance from the center of LG beam. The integers $l_1$ and $l_3$ are the TCs of the respective fields, and $\phi$ denotes the azimuthal angle.
 \begin{figure*}
\centering
\begin{subfigure}{.49\linewidth}
    \includegraphics[width=8cm]{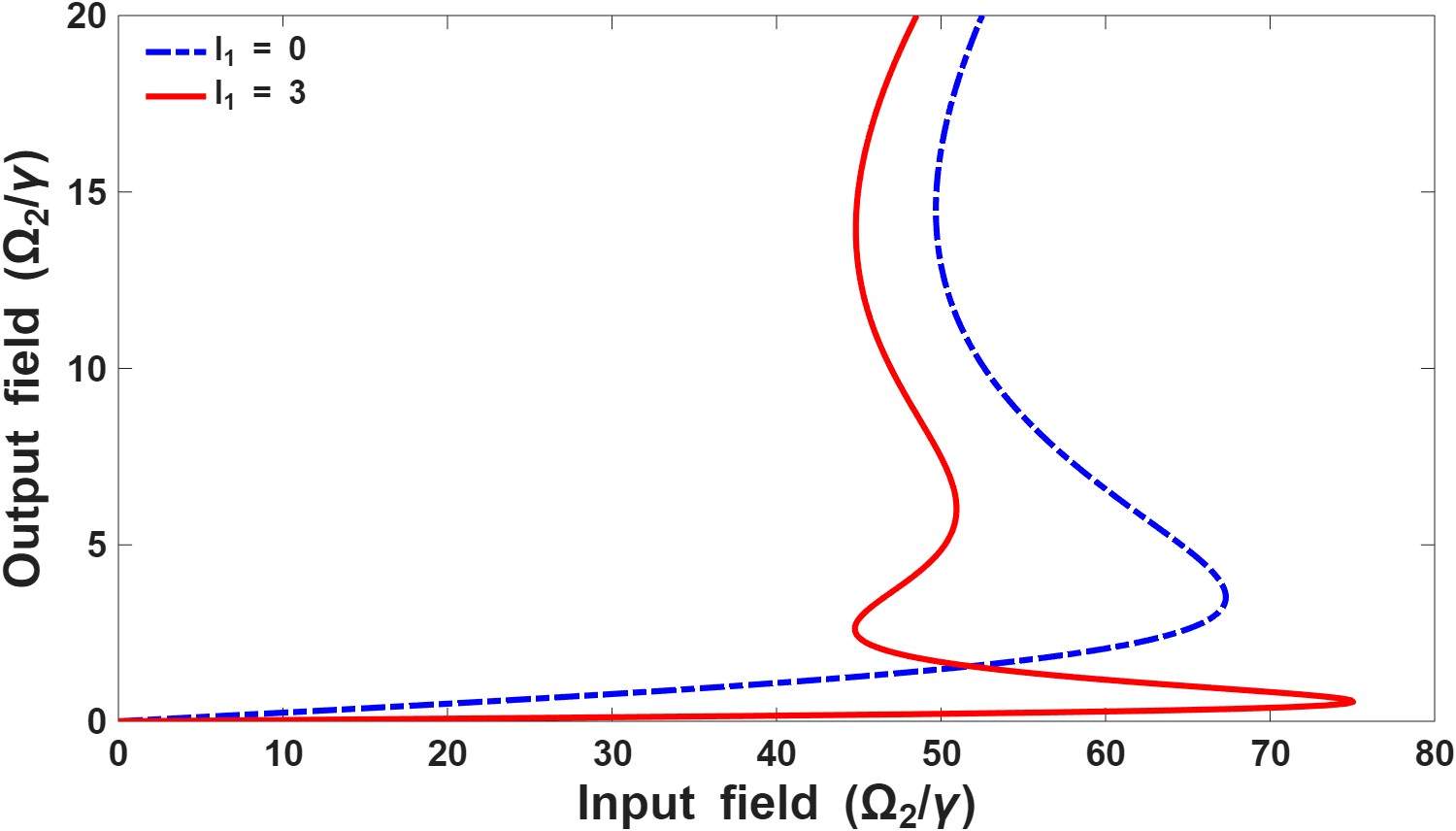}
    \caption{}
    \label{6a}
\end{subfigure}
\hfill
\begin{subfigure}{.49\linewidth}
    \includegraphics[width= 8cm]{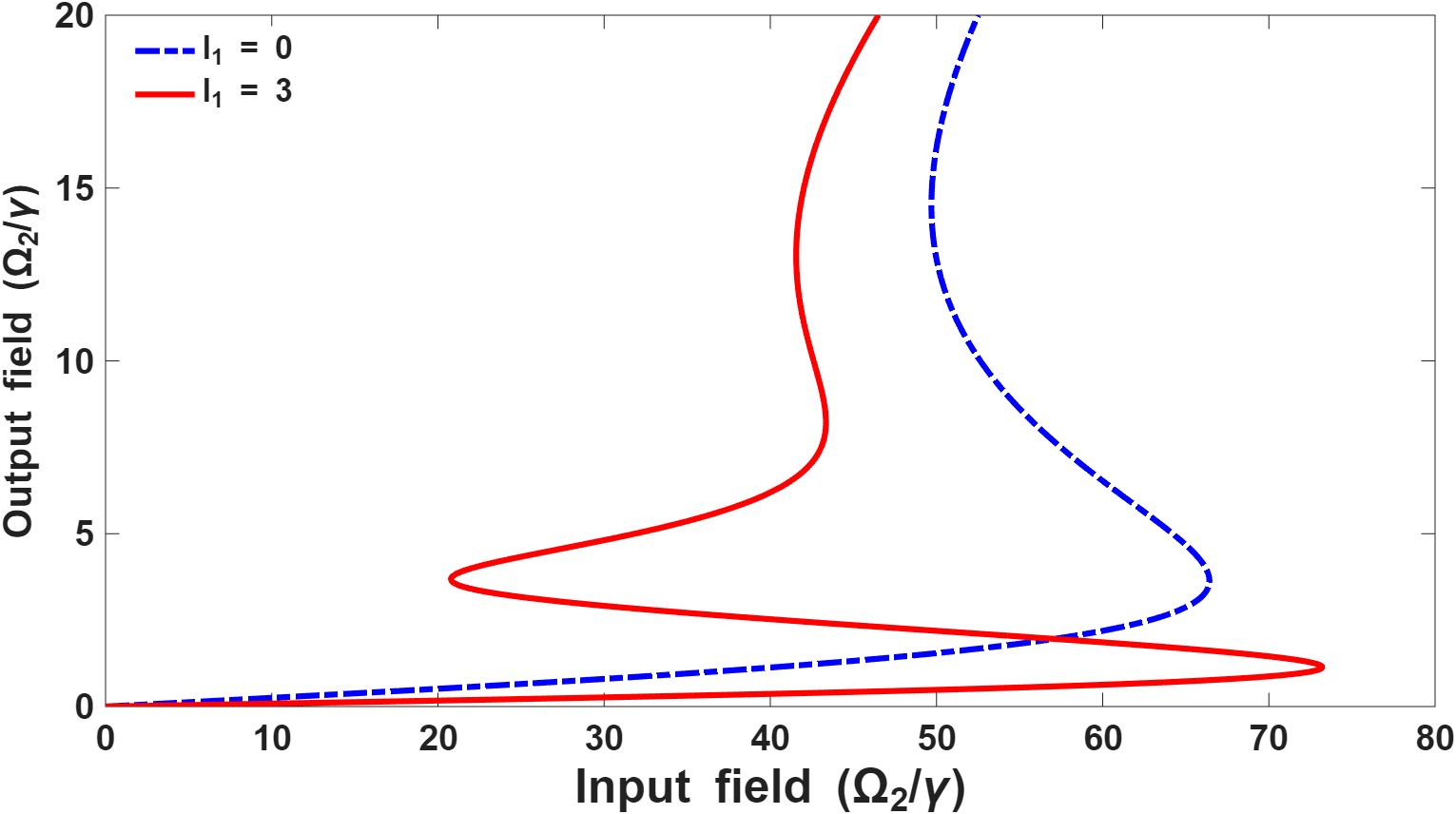}
    \caption{ }
    \label{6b}
\end{subfigure}
\hfill
\begin{subfigure}{.49\linewidth}
    \includegraphics[width= 8cm]{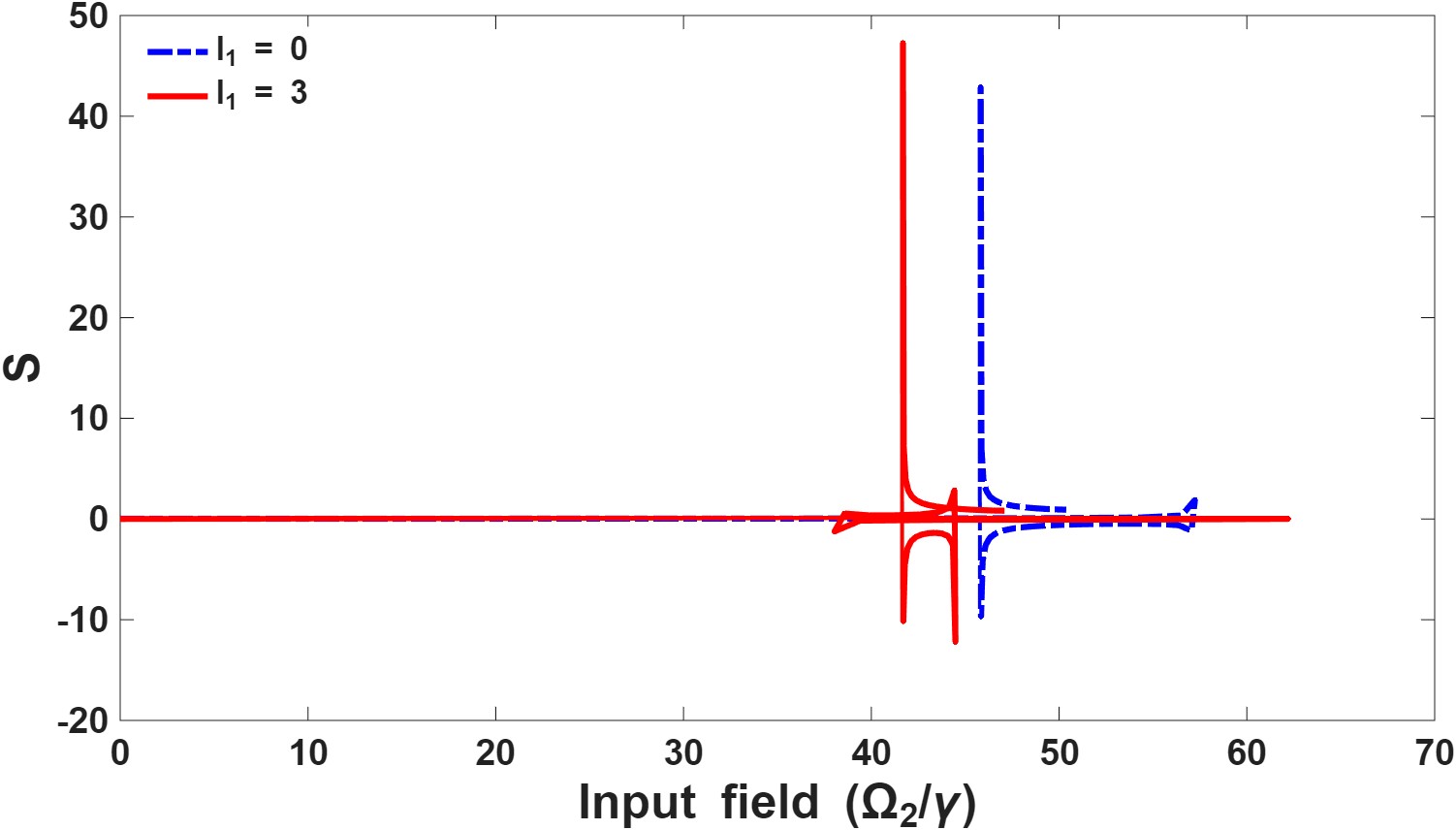}
    \caption{ }
    \label{6c}
\end{subfigure}\hfill
\begin{subfigure}{.49\linewidth}
    \includegraphics[width= 8cm]{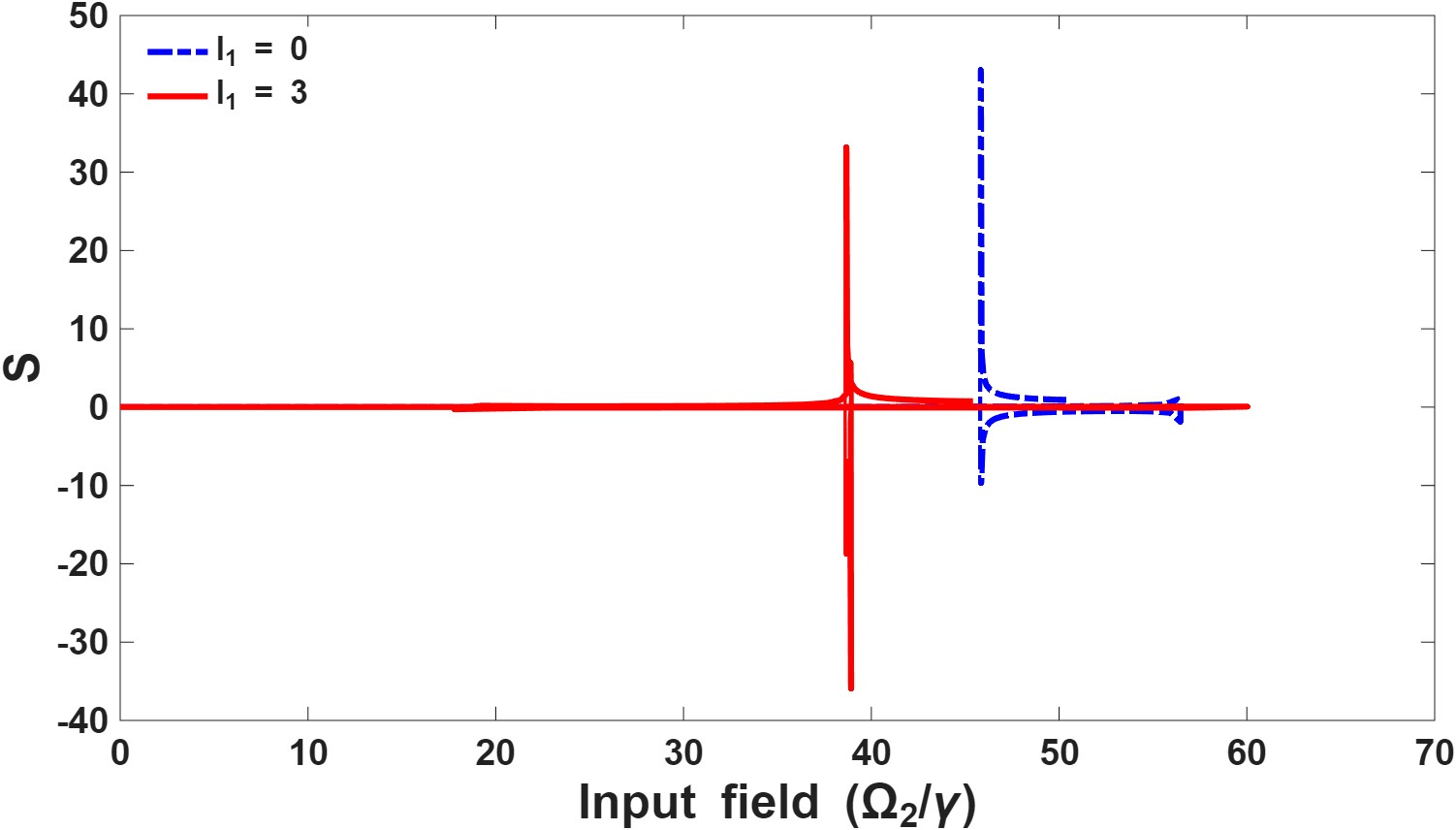}
    \caption{ }
    \label{6d}
\end{subfigure}
\hfill
\caption{
In subfigures (a) and (b), we display the OB behavior for \( l_1 = 0\) (blue, dash-dotted) and 3 (red, solid), for the relative phase (a) \( \phi = \pi/4 \) and (b) \( \phi = \pi/10 \).
In subfigures (c) and (d), we display the slope of OB curves as a function of input Rabi frequency for \( l_1 = 0 \) (blue, dash-dotted) and \( l_1 = 3 \) (red, solid), for (c) \( \phi = \pi/4 \) and (d) \( \phi = \pi/10 \). The other parameters are \( \Delta_1 = \Delta_3 = 0\gamma \), \( \Delta_2 = 7\gamma \), \( C = 300 \), field amplitudes \( A_1 = 5\gamma \), \( A_3 = 0.1\gamma \), and  \( l_3 = 1 \).
 }
\end{figure*}
In Fig.~\ref{6a}, the OB curves are plotted for two different values of the input field’s TC: \( l_1 = 0 \) (blue, dash-dotted) and \( l_1 = 3 \) (red, solid), with the azimuthal angle fixed at \( \phi = \pi/4 \), and \( l_3 = 1 \). The plot shows that increasing the value of \( l_1 \) leads to a larger input threshold and larger unstable regions, indicating enhanced bistability and greater stability of the nonlinear behavior of the atoms.

In Fig.~\ref{6b}, the OB curves are presented under the same TC conditions (\( l_1 = 0 \) and \( l_1 = 3 \)), but with the azimuthal angle reduced to \( \phi = \pi/10 \). The bistability threshold is lower for both cases compared to Fig.~\ref{6a}, and for \( l_1 = 3 \), the OB curve exhibits regions of multistability, suggesting that smaller azimuthal angles further amplify the nonlinear interactions induced by the structured LG beams.

To analyze the switching behavior more precisely, the slope $S$ of the OB curves is examined. In Fig.~\ref{6c}, the slope of the OB response is plotted as a function of input field strength for \( \phi = \pi/4 \), again comparing \( l_1 = 0 \) and \( l_1 = 3 \). The figure clearly shows that the slope is significantly steeper for \( l_1 = 3 \), implying faster and more efficient switching dynamics in the nonlinear regime.

Similarly, in Fig.~\ref{6d}, we show a similar trend of $S$ for \( \phi = \pi/10 \), and the trend persists: the slope is considerably enhanced for the non-zero TC \( l_1 = 3 \), compared to the \( l_1 = 0 \) case. This confirms that the TC plays a critical role not only in modifying the stability of the OB response but also in improving the sensitivity and responsiveness of the switching process. Both the azimuthal phase \( \phi \) and the TC ( $l_1, l_3$) offer tunability to control and enhance the optical bistability in structured-light-mediated atomic systems.

\section{CNOT Gate Based on Optical Bistability}\label{sec:cnot}

Next, we discuss how a CNOT gate can be realized in our system. In this gate, when the control qubit is in the $\ket{1}$ state, the target qubit undergoes a bit-flip operation at the output, whereas no change occurs when the control qubit is in the $\ket{0}$ state. To realize this conditional flipping mechanism, we utilize the phenomenon of OB, as depicted in Fig.~\ref{8a}. In this method, the output state of the target field switches from $\ket{0}$ to $\ket{1}$ or vice versa, depending on its initial state and the presence of optical bistability. In our protocol, we choose $\Omega_1$ as the \textit{control qubit} and $\Omega_2$ as the \textit{target qubit}. The logical states $\ket{0}$ and $\ket{1}$ are encoded in the intensity levels of these fields, with low (or zero) intensity corresponding to $\ket{0}$ and high intensity corresponding to $\ket{1}$. The coupling field $\Omega_3$ is kept switched on, so that the OB feature can be employed to achieve the gate operation. 

\begin{figure*}
\centering
    \includegraphics[width=15cm]{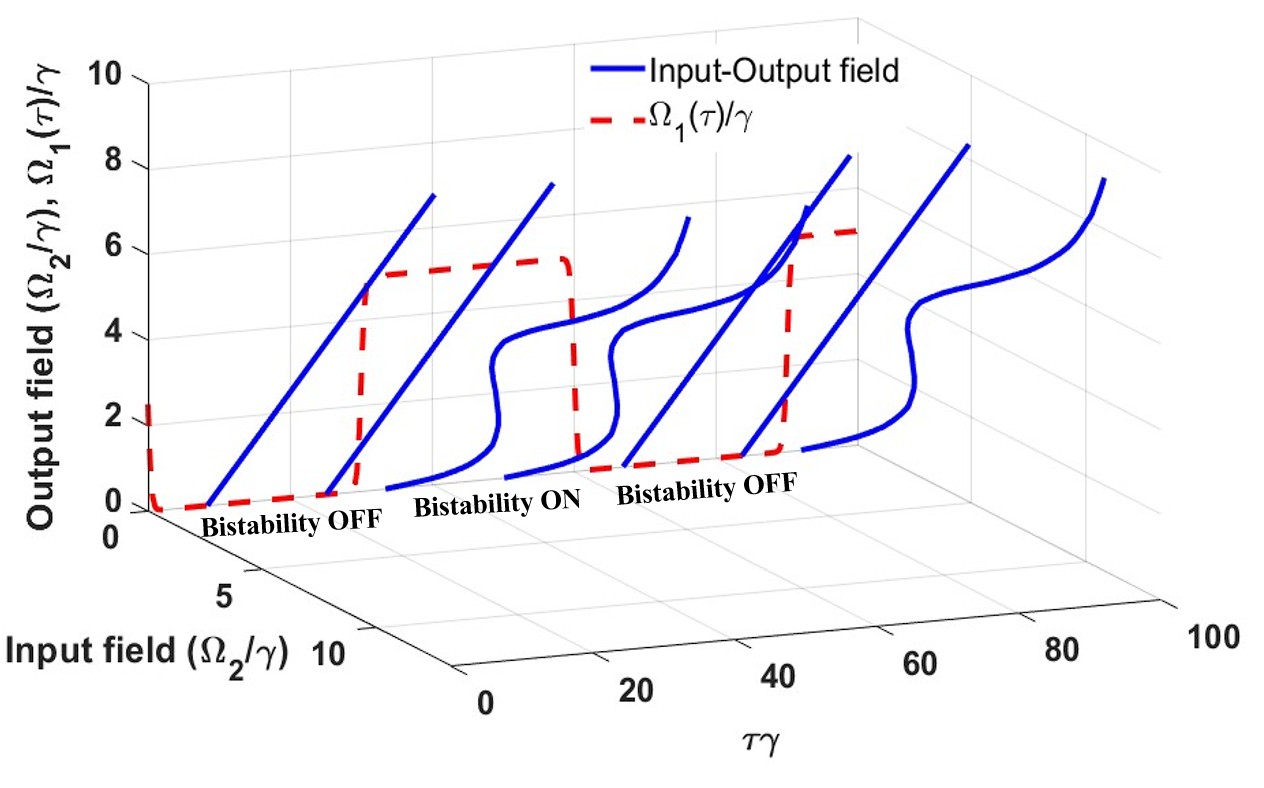}
    \caption{Variation of OB with normalize time \(\tau \gamma\), where the blue solid curves are for the output field vs input field and red dashed curve represents $\Omega_1(\tau)/\gamma)$. The parameters are $\Omega_{01}$=5$\gamma$, $\Omega_3$=5$\gamma$, $\Delta_1$=$\Delta_3$=0$\gamma$, $\Delta_2$=5$\gamma$, and C=300.}
     \label{9}
\end{figure*}

To achieve a dynamic control over the optical bistable response of the system, we implement a time-dependent modulation of the control field \( \Omega_1(\tau) \). Specifically, the control field is designed as a periodic square-like pulse, smoothly shaped using hyperbolic tangent functions to ensure realistic temporal switching. The mathematical form of this control field is given by \cite{Hien2022}:
\begin{equation}
\begin{split}
\Omega_1(\tau) = \Omega_{01} \Big[1 - 0.5\tanh\{2(\tau - 0)\} +  
 0.5\tanh\{2(\tau - 30)\} \\
 - 0.5\tanh\{2(\tau - 60)\} + 0.5\tanh\{2(\tau - 90)\} \Big],
\end{split}
\end{equation}
This function generates a sequence of four consecutive segments in which the control field switches between high and low strength in a smooth, continuous fashion. The amplitude \( \Omega_{01} \) denotes the peak strength of the control field during its ON phase. Each transition from ON to OFF and vice versa occurs over a finite time window, controlled by the argument of the hyperbolic tangent function.
Here we have introduced a coordinate transformation \( \tau = t - \frac{z}{c} \) and \( \zeta = z \). The propagation equation Eq. (\ref{6}), which is a partial differential equation, can then be simplified as an ordinary differential equation as follows, which does not have any explicit spatial dependence in the time evolution
\begin{equation}
      \frac{\partial E_2}{\partial \zeta} = \frac{i\omega_2}{2\epsilon_o}P(\omega_2)  \;.\label{14}
\end{equation}
We further numerically solve Eqs.~(\ref{3}) and (\ref{14}) using the Runge-Kutta method, incorporating the time-dependent control field \( \Omega_1(\tau) \), and apply the boundary conditions specified in Eq.~\ref{7}. The simulation results are shown in Fig.~\ref{9}, where the solid blue curves represent the input-output field strength characteristics as a function of normalized time $\tau\gamma$, and the dashed red line illustrates the temporal profile of the control field \( \Omega_1(\tau)/\gamma \).

As depicted in Fig.~\ref{9}, the bistable behavior of the system can be dynamically modulated by varying the control field \( \Omega_1(\tau) \) in time. During intervals when the control field is high [\( \Omega_1(\tau) \approx \Omega_{01} = 5\gamma \)], the system displays a pronounced bistable response, characterized by an S-shaped input-output curve. In this regime, small changes in input field strength induce large, abrupt shifts in output field, signifying the switching behavior necessary for realizing a logical NOT operation on the probe (target) field when the control input is in the logical state \( \ket{1} \).

Conversely, during time intervals where the control field is suppressed [\( \Omega_1(\tau) \approx 0 \)], the bistability vanishes, and the system exhibits a linear or monostable response. This corresponds to the logical state \( \ket{0} \) of the control input, wherein the target output remains unchanged. Such behavior satisfies the conditional nature of a CNOT gate: the target field only switches when the control field is active.

Therefore, the time-dependent field \( \Omega_1(\tau) \) effectively acts as the logical control input, toggling the system between bistable (logic-flipping) and monostable (no-flip) regimes. This dynamic switching mechanism realizes the core logical condition of a CNOT gate: the probe (target) field undergoes switching if and only if the control field is high (\( \ket{1} \)), and remains unaffected when the control field is low (\( \ket{0} \)).


This behavior confirms that the system successfully implements the logic operations of a CNOT gate. Furthermore, the stability and tunability of the optical logic gate can be precisely engineered by adjusting physical parameters such as the Rabi frequencies and detunings of the fields, the cooperation parameter \( C \), and the angular momentum index \( l \) of the LG beams, as detailed in Sec. \ref{num ob} and Sec.
\ref{num ob}. As shown in Sec. \ref{oam ob}, under suitable conditions, the system can exhibit two bistability regions, offering even greater flexibility for logic encoding and switching thresholds.

\subsection{Performance analysis of the gate operation}
To quantify the reliability of logical state identification in our bistable system, we evaluate the intra-branch variation in output field strength for each logic state. Specifically, the logical states \(|0\rangle\) and \(|1\rangle\) are represented by the lower and upper branches of the hysteresis curve, as shown in Fig.\ref{8a}, respectively. In an ideal scenario, each logic state would correspond to a single, well-defined output field value. However, due to the finite slope of the bistable branches and nonlinear system response, the output field strength for each state spans a finite range. This variation introduces ambiguity in interpreting the logic level from the output signal.

We define the percentage error for each state as the normalized average deviation of field strength values within that branch from their mean. A smaller percentage error indicates that the output field strengths are more tightly clustered around a single representative value, thereby reducing the overlap between the distributions of \(|0\rangle\) and \(|1\rangle\) states. Consequently, the distinguishability of the logic levels improves, leading to higher fidelity in logic operations. Hence, the percentage error serves as a practical proxy for evaluating the accuracy and performance of bistable logic gates based on the output field.

\begin{figure*}[ht]
\centering
\begin{subfigure}{.5\linewidth}
    \includegraphics[width=8cm]{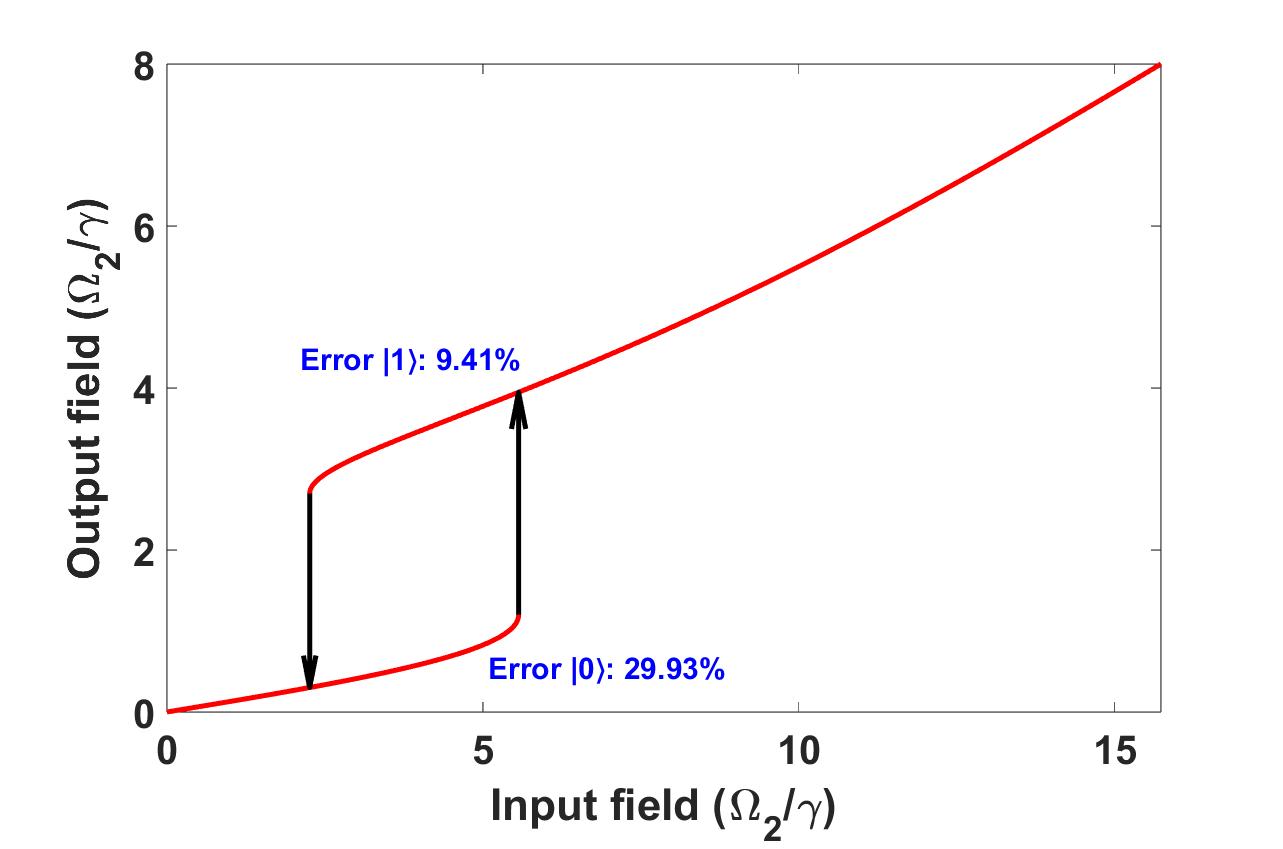}
    \caption{}
    \label{9a}
\end{subfigure}
\hfill
\begin{subfigure}{.49\linewidth}
    \includegraphics[width=8cm]{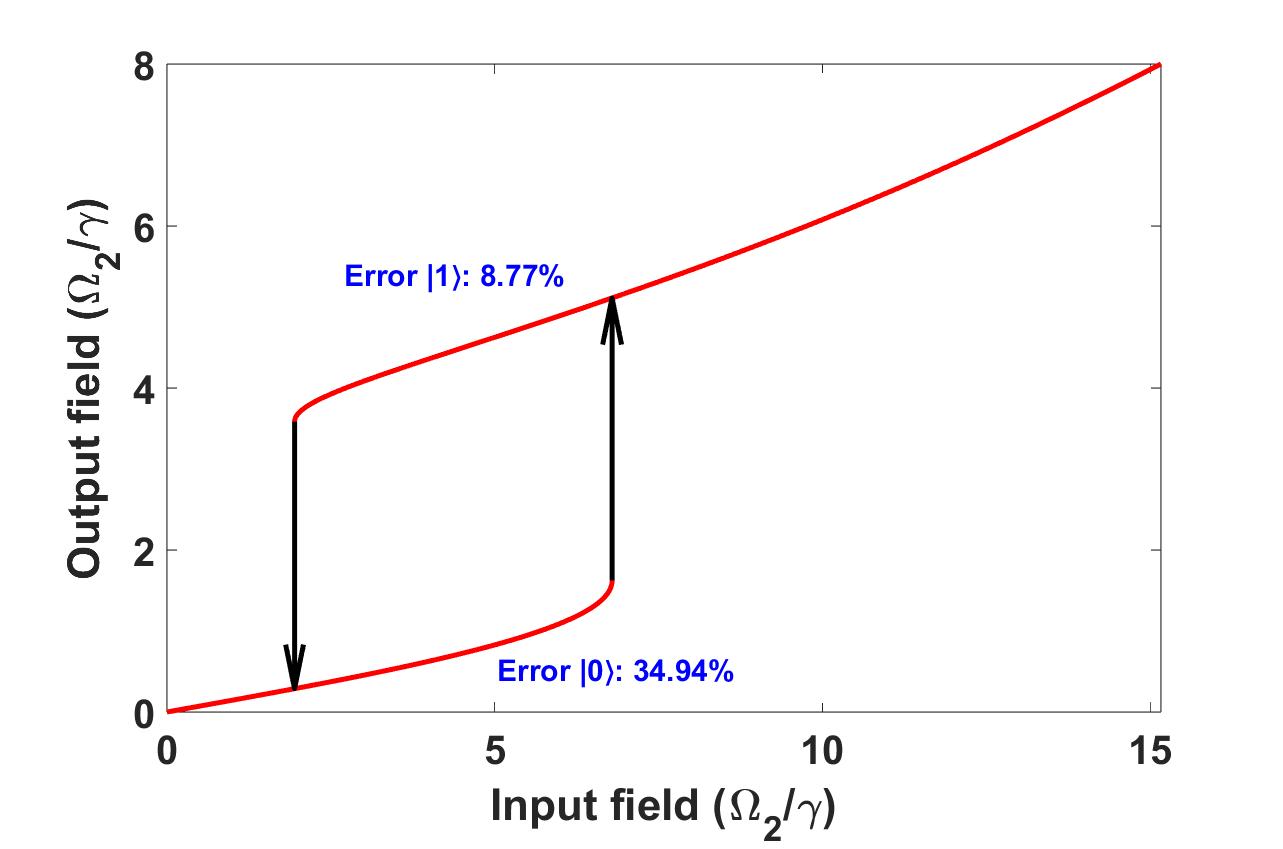}
    \caption{}
    \label{9b}
\end{subfigure}
\hfill
\begin{subfigure}{.5\linewidth}
    \includegraphics[width=8cm]{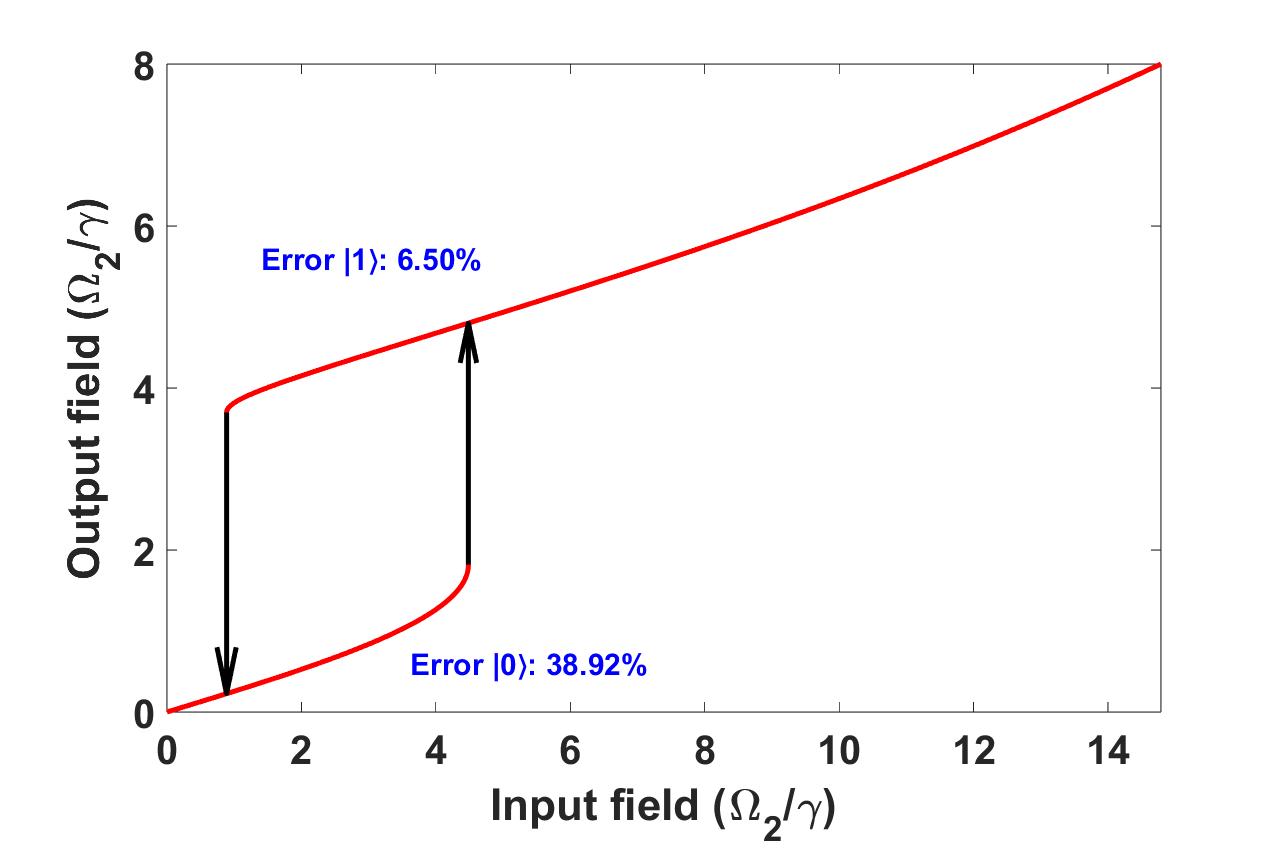}
    \caption{}
    \label{9c}
\end{subfigure}
\hfill
\begin{subfigure}{.49\linewidth}
    \includegraphics[width=8cm]{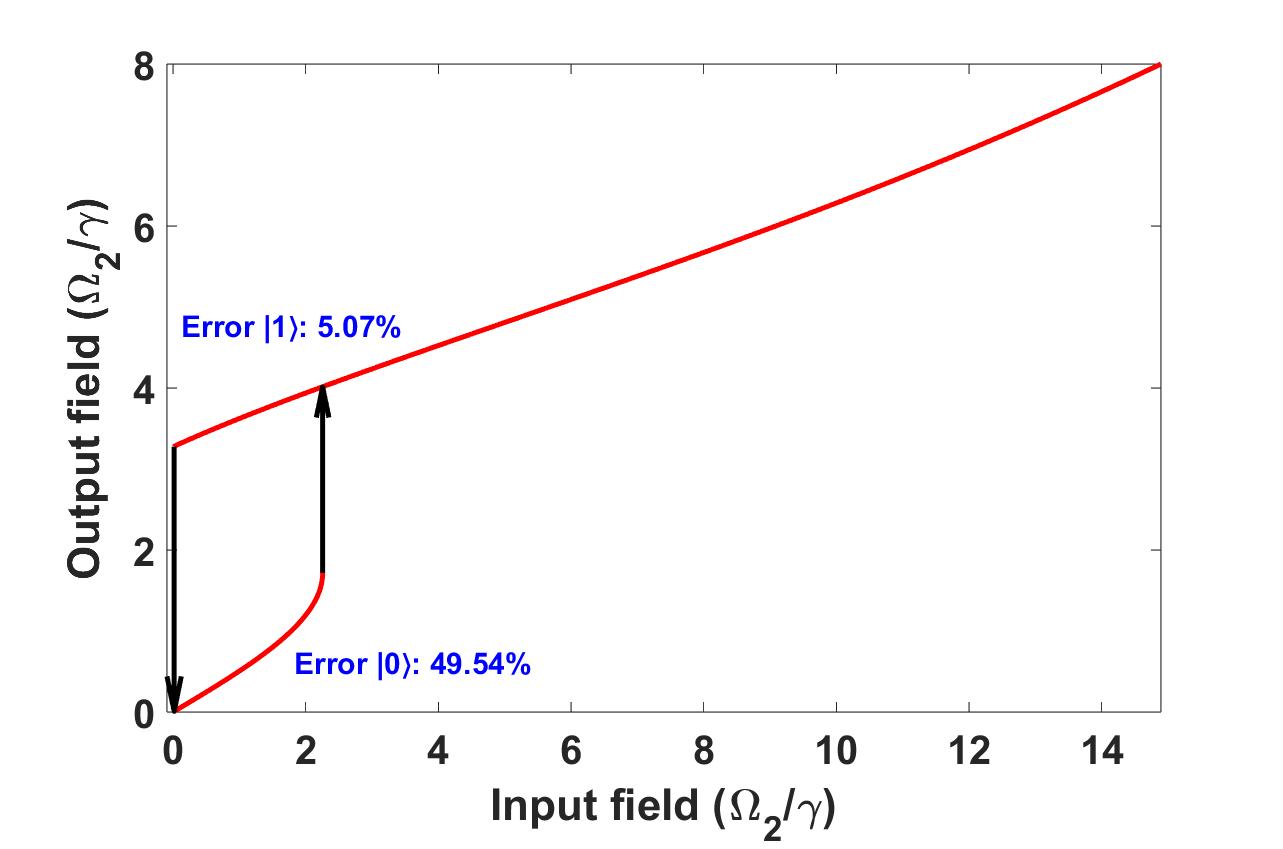}
    \caption{}
    \label{9d}
\end{subfigure}
\caption{Hysteresis curves for different values of $\Omega_1$: (a) $\gamma$, (b) $2\gamma$, (c) $3\gamma$, and (d) $4\gamma$. The percentage errors (blue) associated with logic states $\ket{1}$ and $\ket{0}$ quantify the intensity variation within the upper and lower branches of the bistability curve, respectively. All other system parameters are identical to those used in Fig. 8.}
\end{figure*}

We show in Figs.~\ref{9a}–\ref{9d} the behaviour of the bistable output field as a function of input field for four different values of \(\Omega_1\), with the probe detuning fixed at \(\Delta_2 = 5\gamma\) and \(\Omega_3 = 5\gamma\). In each subplot, forward and backward scans of the input field reveal the characteristic hysteresis behavior of the system. The black vertical arrows indicate the switching transitions between logic states \(|0\rangle\) and \(|1\rangle\), and the blue annotations mark the calculated percentage errors for each state.

It is observed that increasing \(\Omega_1\) leads to a slowly progressive steepening of the upper branch, which reduces intensity fluctuations and thereby lowers the percentage error for logic state \(|1\rangle\). In contrast, the lower branch associated with logic state \(|0\rangle\) does not show a uniform improvement; its error either remains significant or increases, particularly at higher values of \(\Omega_1\). This asymmetric error behavior highlights the greater susceptibility of the lower branch to slope-induced variations, which may limit the precision and reliability of the logic \(|0\rangle\) state.

\begin{figure*}[ht]
\centering
\begin{subfigure}{.5\linewidth}
    \includegraphics[width=8cm]{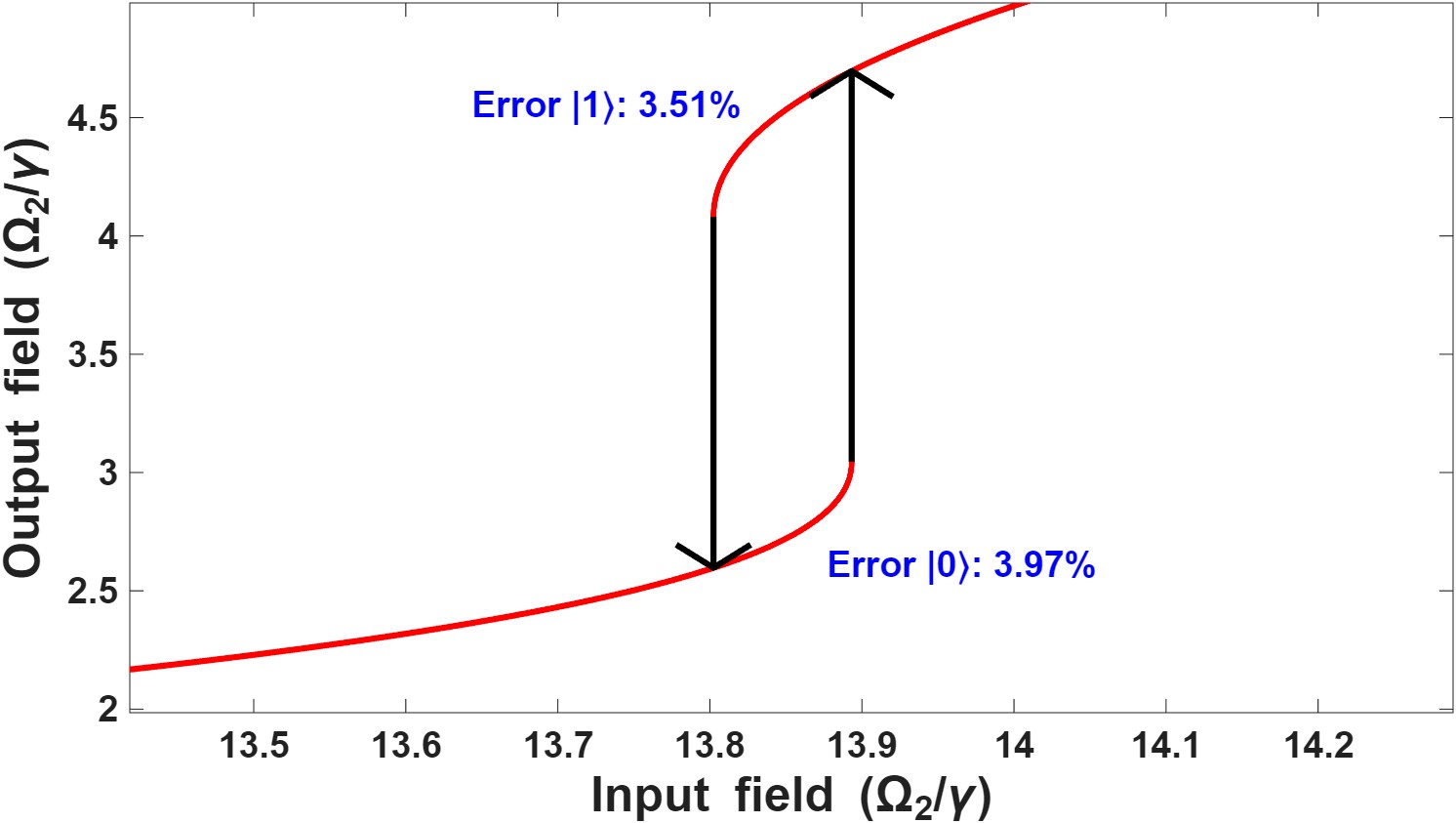}
    \caption{}
    \label{10a}
\end{subfigure}
\hfill
\begin{subfigure}{.49\linewidth}
    \includegraphics[width=8cm]{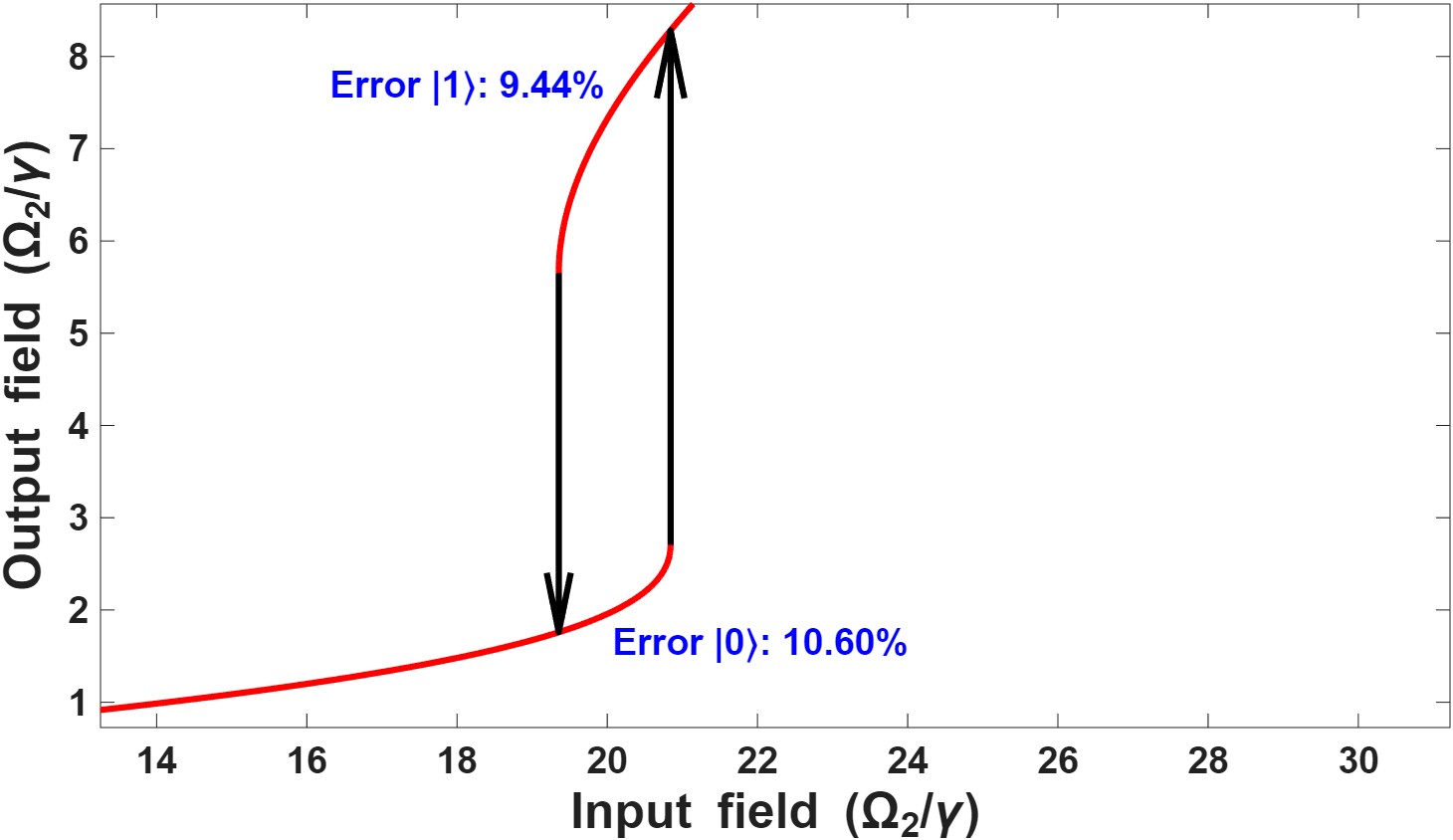}
    \caption{}
    \label{10b}
\end{subfigure}
\caption{Hysteresis curves corresponding to different detuning values: (a) \(\Delta_1 = 8\gamma\) and (b) \(\Delta_3 = 6\gamma\). The percentage errors (blue) associated with the logical states \(\ket{1}\) and \(\ket{0}\) quantify the intensity fluctuations observed within the upper and lower branches of the bistability curves, respectively. All other system parameters are the same as those used in Fig.~8.
}
\end{figure*}

To further explore the impact of system parameters on state fidelity, we fix \(\Omega_1 = \Omega_3 = 5\gamma\) and vary the detunings \(\Delta_1\) and \(\Delta_3\). When \(\Delta_1 = 8\gamma\), the percentage error for logic state \(|1\rangle\) is reduced to \(3.51\%\), while that for logic state \(|0\rangle\) is \(3.99\%\). Similarly, for \(\Delta_3 = 6\gamma\), the error in logic state \(|1\rangle\) is \(9.44\%\), and in state \(|0\rangle\), it is \(10.60\%\). These values represent a significant improvement over the results shown in Figs.~\ref{10a}–\ref{10b}, indicating that appropriate detuning can effectively suppress intra-branch fluctuations and enhance logic-state fidelity.

\section{Experimental Realization}\label{sec:experiment}
The proposed theoretical model can be experimentally implemented using a cold \({}^{87}\mathrm{Rb}\) atomic ensemble on the \(D_1\) transition line. Cold atomic systems, particularly those based on laser-cooled and magneto-optically trapped rubidium atoms, offer several advantages: they suppress Doppler broadening, provide high optical depth, and maintain long coherence times, thereby enabling precise control over atomic interactions and quantum optical phenomena.  

A high optical depth, essential for strong light-matter coupling, can be achieved in cold rubidium ensembles confined to a small interaction volume, where atomic densities on the order of \(10^9\) atoms/cm\(^3\) are routinely attainable. Owing to their high degree of isolation and controllability, such systems are well-suited for investigating nonlinear optical effects, including optical bistability~\cite{PhysRevA.70.023802,PhysRevA.91.013820} and the CNOT gate. Furthermore, the interaction of LG beams with the cold atomic ensemble introduces an additional degree of spatial and phase control, making the proposed scheme a strong candidate for practical quantum computing and optical information processing applications.  

The specific energy-level configuration corresponds to the \({}^{87}\mathrm{Rb}\) \(D_1\) (5\(^2S_{1/2} \rightarrow 5^2P_{1/2})\) transition, with states defined as \(\ket{1} \equiv \ket{5^2S_{1/2}, F=1}\), \(\ket{2} \equiv \ket{5^2S_{1/2}, F=2}\), \(\ket{3} \equiv \ket{5^2P_{1/2}, F'=1}\), and \(\ket{4} \equiv \ket{5^2P_{1/2}, F'=2}\). The spontaneous decay rates are \(\gamma_{13} = 0.25\Gamma\), \(\gamma_{23} = 0.75\Gamma\), \(\gamma_{41} = 0.625\Gamma\), and \(\gamma_{42} = 0.375\Gamma\), respectively, where \(\Gamma = 2\pi \times 5.75\)~MHz denotes the natural linewidth~\cite{PhysRevA.70.023802,PhysRevA.91.013820,steck2003rubidium}.  

In this configuration, the presence of a strong coupling field \(\Omega_3\) on the \(\ket{2} \leftrightarrow \ket{4}\) transition transforms the system into a gain medium for the weak probe field \(\Omega_2\). This gain arises from the nonzero decay channel \(\gamma_{14}\), which enables population recycling and establishes a constructive quantum interference pathway in the population flow, thereby amplifying the probe. In contrast, when \(\Omega_3 = 0\), this interference pathway is absent, and the system reduces to a conventional EIT configuration for the probe field.  


\section{Conclusion}\label{sec:conclusion}
This work presents a comprehensive investigation of the probe field dynamics in a multi-level atomic system, with particular focus on achieving coherent control over its absorption and gain characteristics. Through systematic analysis, we explore the conditions required for realizing OB, emphasizing control over key performance metrics such as threshold behavior, stability, and switching efficiency. Our results reveal that several physical parameters, including field detunings, atomic density, and Rabi frequencies, play a pivotal role in shaping the OB response. Notably, the introduction of gain for the probe field significantly enhances the system’s nonlinear interaction, enabling bistable behavior to emerge at reduced input intensities due to lowered switching thresholds.

To further augment the tunability of the system, we incorporate structured light beams with OAM, thereby introducing an additional degree of control over the OB characteristics. Moreover, we demonstrate that one of the input fields can serve as an effective dynamic control knob, capable of modulating the bistable response of a second field. This mutual coupling introduces a versatile and reconfigurable mechanism for all-optical switching.

Crucially, we show that under suitable conditions, the system satisfies the essential logical criteria for implementing a CNOT gate. The logical states \(|0\rangle\) and \(|1\rangle\) are represented by output intensity levels corresponding to distinct branches of the hysteresis loop, and we quantify their reliability by calculating the percentage error in intra-branch field strength variation. Our findings underscore the critical importance of optimizing system parameters to minimize intra-branch field strength uncertainty in optical bistability. In particular, the reduction of field strength variation in the lower bistable branch is essential for enhancing the fidelity and stability of optical logic operations. Such precision is crucial for the practical realization of high-performance bistability-based quantum logic gates, which are well-suited for integration into photonic and quantum information processing architectures.


Moreover, building upon the controllable and highly tunable bistable behavior demonstrated in this system, the extension toward more complex logic operations, such as the implementation of a Toffoli gate (also called a \textit{ controlled-controlled-NOT} or CCNOT gate) is both feasible and promising within the same optical framework. By adopting a five-level atomic configuration, the system can support two independently addressable control fields. These fields act as logical control qubits whose simultaneous high-intensity states are required to induce bistable switching in the probe field. As a result, the probe field exhibits bistability only when both control fields are active, thereby satisfying the logical conditions required for a Toffoli gate. This approach enables the realization of a fully optical Toffoli gate, expanding the utility of the system beyond two-input logic operations. It demonstrates the versatility and scalability of the bistable atomic platform in enabling more sophisticated quantum and classical optical computing functionalities. With precise parameter control and field configuration, such systems pave the way for constructing integrated all-optical circuits capable of implementing complex logic schemes within compact, tunable atomic architectures.

\section{Acknowledgment}
The authors acknowledge the financial support provided by the Council for Scientific and Industrial Research (CSIR), India, during this work.

\bibliography{apssamp}

\end{document}